\documentclass[epj]{svjour}
\usepackage[english]{babel}
\usepackage{graphicx}
\renewcommand{\textfraction}{.0}

\begin{document}

\title{Detailed study of the decay of $^{32}$Ar} 

\author
{B.~Blank\inst{1} \and
N. Adimi\inst{2} \and
M. Alcorta\inst{3} \and
A. Bey\inst{1} \and
M.J.G Borge\inst{3} \and
B.A. Brown\inst{4} \and
F. de Oliveira Santos\inst{5} \and 
C. Dossat\inst{1} \and             
H.O.U. Fynbo\inst{6} \and          
J. Giovinazzo\inst{1} \and         
H.H. Knudsen\inst{6} \and          
M. Madurga\inst{3} \and    
A. Magilligan\inst{4} \and         
I. Matea\inst{7} \and              
A. Perea\inst{3} \and              
K. S{\"u}mmerer\inst{8} \and       
O. Tengblad\inst{3} \and           
J.C. Thomas\inst{5}    
}

\institute{Centre d'Etudes Nucl\'eaires de Bordeaux Gradignan,
UMR 5797 CNRS/IN2P3 - Universit\'e de Bordeaux, 19 Chemin du Solarium, CS 10120, F-33175 Gradignan Cedex, France  \and
Laboratoire SNIRM, Facult\'e de Physique, USTHB, B.P.32, El Alia, 16111 Bab Ezzouar, Alger, Algeria \and
Instituto de Estructura de la Materia, CSIC, Serrano 113bis, E-28006-Madrid, Spain  \and
Department of Physics and Astronomy, and National Superconducting Cyclotron Laboratory, 
Michigan State University, East Lansing, Michigan 48824-1321, USA \and
Grand Acc\'el\'erateur National d'Ions Lourds, CEA/DRF-CNRS/IN2P3, B.P. 55027, F-14076 Caen Cedex 05, France \and
Department of Physics and Astronomy, University of Aarhus, Ny Munkegade 1520, DK-8000 Aarhus C, Denmark \and
Laboratoire de Physique des 2 Infinis Ir{\`e}ne Joliot-Curie - IJCLab, B{\^a}t. 100, 15 rue Georges Cl{\'e}menceau, 91405 Orsay Cedex, France \and
Gesellschaft f\"ur Schwerionenforschung mbH, Planckstrasse 1, D-64291 Darmstadt, Germany}

\renewcommand{\textfraction}{.01}
\headsep 1.7cm

\abstract{
In an experiment performed at the SPIRAL1 facility of GANIL, the $\beta$ decay of $^{32}$Ar has been studied by means of the "Silicon Cube" device
associated with germanium clover detectors from the EXO\-GAM array. Beta-delayed protons and $\gamma$ rays have been observed and allowed the determination
of all relevant decay branches. The Gamow-Teller strength distribution is compared to shell-model calculations and excellent agreement is found. The Fermi
strength is inline with expectations. A quasi-complete decay scheme of $^{32}$Ar is established.
}

\authorrunning{B. Blank {\it et al.}}
\titlerunning{Detailed study of the decay of $^{32}$Ar}

\maketitle

\section{Introduction}

Nuclear $\beta$ decay is a powerful tool for the study of nuclear structure for nuclei far from stability. It is a high-precision tool that allows the
extraction of effects linked e.g. to the pairing interaction, nuclear deformation, and questions related to the exact formulation of the weak interaction.
Beyond its importance for nuclear structure and fundamental interactions, nuclear $\beta$ decay plays also an important role for basically
all types of stars at any moment of their evolution. The access to nuclear decay information is facilitated by the fact that $\beta$-decay experiments
are among the most simple-to-implement experiments in nuclear physics. Therefore, beyond the observation of a new nuclide, the first investigations are usually 
carried out by studying its decay properties~\cite{blank08review}.

\begin{table*}[hht]
\begin{center}
\caption{The thicknesses of the different silicon detectors used in the present study are given. The arrangement of the detectors is presented
         in figure~\ref{fig:setup}.}
\begin{tabular}{lcccccc}
\hline
\hline
\rule{0pt}{1.3em}
Detectors &  1  &  2  &  3  &  4  &  5  &  6  \\
[0.5em]\hline
\rule{0pt}{1.3em}
DSSSD thickness ($\mu$m) & 300 & 287 & 270 &  64	& 1000 & 288 \\
Large-area detector thickness ($\mu$m) & 300 & 300 & 500 & 1473& 150 & 1498 \\
[0.5em]\hline 
\hline
\end{tabular}
\end{center}
\label{tab:thickness}
\end{table*}

The present paper describes a study of the $\beta$-decay properties of $^{32}$Ar. This nucleus is perhaps the most pro\-ton-rich isotope of argon bound by the strong 
interaction, as $^{31}$Ar, its neighbour, is predicted by some mass models to be (close to) proton unbound~\cite{pape88,comay88a,duflo95,moeller16}. This nucleus 
is also a candidate for precision studies of beyond-standard-model contributions to the weak interaction via the search for scaler and tensor 
currents~\cite{schardt93,adelberger99,araujo20}. Its $\beta$-decay properties have been studied four times in experiments, three times at ISOLDE~\cite{schardt93,hagberg77,bjornstad85} 
and a fourth time in two experiments at NSCL and at ISOLDE~\cite{bhattacharya08}. While the first experiment of Hagberg {\it et al.}~\cite{hagberg77} only observed protons from the isobaric analogue state (IAS) 
and verified the isobaric multiplet mass equation (IMME), the second experiment of Bj{\"o}rnstad {\it et al.}~\cite{bjornstad85} was mainly concerned with the giant Gamow-Teller (GT) 
resonance. The third work of Schardt and Riisager~\cite{schardt93} extracted limits of exotic currents of the weak interaction. The last work executed by Bhattacharya 
{\it et al.}~\cite{bhattacharya08} dealt with the Fermi strength to determine as precisely as possible the decay strength, the $ft$ value, for the super-allowed $\beta$ decay 
of $^{32}$Ar in order to deduce the isospin-impurity correction $\delta_C$ experimentally and compare it to a theoretical prediction. Unfortunately, the authors of this last work 
were not interested in extracting the distribution of the GT strength, which allows a relatively precise comparison with shell-model calculations. The main drawback for the rather 
detailed data of Bj{\"o}rnstad {\it et al.}~\cite{bjornstad85} is probably the very low $\gamma$-ray detection efficiency, which did not allow the authors to attribute all proton 
groups observed correctly to decays to the first excited state in $^{31}$S (see below). We will compare our results only to the data of Bj{\"o}rnstad {\it et al.} and Bhattacharya 
{\it et  al.}, because Schardt {\it et al.} gave information only for a few proton lines and Hagberg {\it et al.} observed only the decay of the IAS.

The experiment described here was performed at the identification station of SPIRAL1 at GANIL. The data were taken at the same time as data for $^{31,33}$Ar 
already published~\cite{matea09,adimi10}. For the purpose of the present experiment, part of the identification station was dismounted and replaced by the "Silicon Cube" 
array~\cite{matea09}. Although we will also investigate the Fermi strength in the decay of $^{32}$Ar, the focus of the paper is on the GT strength distribution, 
which we will compare to predictions of the nuclear shell model.

\section{Experimental procedure and set-up}

The $^{32}$Ar$^{3+}$ beam was produced by projectile fragmentation of a $^{36}$Ar primary beam at 95~MeV/A accelerated by the CSS cyclotrons of GANIL,
which impinged on the SPIRAL1 graphite target. The $^{32}$Ar atoms diffusing out of the target and reaching the NANOGAN-III ECR ion source of SPIRAL1 were 
subsequently ionised and sent to the SPIRAL1 identification station with an energy of 30~keV, where they were intercepted by a 0.9$\mu$m thick mylar foil. 
The beam line optics between the target-ion source and the
detection set-up was regularly optimised by means of a stable $^{40}$Ar beam also produced by the ion source.

\begin{figure}[hht]
\begin{centering}
\includegraphics[scale=0.5,angle=0]{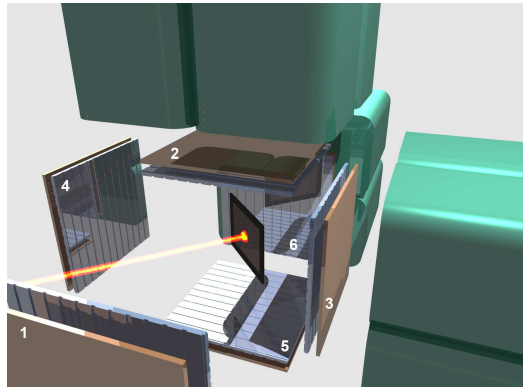}
\par\end{centering}
\caption{{\small{Experimental set-up of the present experiment. Six double-sided silicon strip detectors form a cube. They are backed by large-area 
                 silicon detectors. Three EXOGAM clover detectors complete the set-up. The activity is deposited on a catcher foil in the center of 
                 the set-up.}}}
\label{fig:setup}         
\end{figure}

The experimental set-up consisted of a cube of six double-sided silicon strip detectors (DSSSDs, 16 X and 16 Y strips with a width of 3~mm) 
backed by large-area silicon detectors (LASDs, 50~mm x 50~mm). The different thicknesses of the detectors are given in table~\ref{tab:thickness} and 
their position with respect to the incoming beam is shown in figure~\ref{fig:setup}.

All DSSSD channels were read-out by means of 16-channel pre-amplifier cards mounted directly on the vacuum chamber of the detection set-up
and connected to the detectors via a printed circuit board. The LASDs were connected to their pre-amplifiers by means of LEMO cables. 
The total detection efficiency of this set-up can reach up to 54\%~\cite{matea09}, if all detectors are fully operational. In the present paper, we used for the
main analysis only half of the detectors (see in the following paragraph).
The Silicon Cube was surrounded by three EXOGAM clover detectors.

Detectors DSSSD 1, 3, 6 worked without major problems (one Y strip missing for DSSSD~3, DSSSD~5, and DSSSD~6 ) and were used for the main data analysis, 
except for the highest proton energies (see below).
Technical problems, which we could not solve prior to the data taking, prevented DSSSD 2 from functioning correctly. It was therefore excluded
from the data analysis. The thickness of detector 4 was initially chosen to be thin for the study of the $\beta$-delayed 2p decay of $^{31}$Ar. 
For the mainly high-energy protons in the present study, it was too thin and also excluded from the data analysis. DSSSD~5 has, due to its thickness, 
a large contribution to the spectrum from $\beta$ particles. This is not a problem for high-energy protons, but "pollutes" the spectrum at low proton
energies. However, protons with energies above 6000~keV punch through the $\approx$300$\mu$m silicon detectors. Therefore, we used DSSSD~5 for protons 
above 5700~keV (proton groups 29-31 in table~\ref{tab:protons}), instead of detectors 1, 3, and 6.

The large-area detectors, which were meant to be used in anti-coincidence with the DSSSDs in front of them, were calibrated with $\alpha$ particles prior to the data taking. However, 
we did not use them in the present analysis, because the probability of having the $\beta$ particle and the proton in the same pixel of a DSSSD (see below) is so small that this effect 
was negligible. We did not use them either to detect high-energy protons punching through the 300$\mu$m DSSSD, because of their unknown dead-layer and the fact that only an $\alpha$-particle
calibration was available.

The energy calibration of the DSSSDs was performed with well-known energies from the literature~\cite{nndcA=32}.
Precise laboratory energies of 14 proton groups from 1210~keV to 6483~keV were calculated from known excited states in either $^{32}$Cl or $^{33}$Cl and used to calibrate 
the proton spectra taken for the decays of $^{32}$Ar and $^{33}$Ar. The precision of the energy of these calibration proton lines ranges from 0.4~keV for the lowest proton
energies to 8~keV for the highest energies. As it is not possible to determine an energy uncertainty from the error propagation based on 128 different strips
for the four DSSSDs used in the present analysis, we determined the difference between the expected energy and the one determined in the fit of the
experimental spectra for four proton groups used in the calibration. We found an RMS deviation of 4~keV, which we have added quadratically to all proton energy uncertainties
given in the present work. 

Absolute proton detection efficiencies are not needed, because all proton emission branching ratios will be normalised with
respect to the most intense proton group from the IAS. The only assumption is that the detection efficiency is the same for all proton
energies, which is a good assumption, because the proton detection efficiency should be the solid angle covered by the detectors used with respect to 4$\pi$.

The energy resolution obtained for the DSSSD sum spectrum from detectors 1, 3, and 6 was of the order of 50~keV (FWHM).
To accept an event the signals in the front and back side had to be within $\pm$ 125~keV of each other. Once this condition verified, the energy was taken from 
the better of the two sides of each DSSSD (Y side for DSSSDs 1 and 6, X side for the others).

The EXOGAM clover detectors were calibrated in efficiency with the standard calibration sources $^{60}$Co, $^{133}$Ba, $^{137}$Cs, and $^{207}$Bi.
Each of the four elements of the clover detectors was calibrated independently and the spectra were summed after calibration. One segment of clover 1 had a 
drifting gain and was removed from the data analysis. No add-back was used. 
The efficiency curve was fit with a straight line on a double-logarithmic scale. We found an efficiency of 3.29(25)\% at 1~MeV. This
efficiency curve is correct down to energies of 150 - 200 keV, where the entrance window thickness of the germanium detectors and other effects start playing 
a role and where the straight line on a double-logarithmic scale is no longer valid. The 89.9~keV line from the decay of $^{32}$Ar was 
cut by the ADC threshold in some of the germanium crystals (see below).

The energy calibration was performed crystal by crystal with 11 on-line $\gamma$-ray lines from 511 keV to 4772 keV. The precision of the calibration
was checked with the sum spectrum from the 11 segments out of 12 kept for the analysis. A RMS deviation of 0.4~keV was found for the 11 calibration lines 
and added quadratically to the energy uncertainty for all $\gamma$ rays from the decay of $^{32}$Ar and of its daughter nuclei. Due to the fact that the 
89.9~keV line was partially cut, we refrain from giving an energy value for the present experiment.

For the present analysis, the only data acquisition triggers used were triggers from the DSSSDs. Therefore, the trigger efficiency is different for $\beta$-delayed
proton events (higher efficiency) and for $\beta$-delayed $\gamma$ rays (lower trigger efficiency).

Runs on $^{31,32,33,34}$Ar were performed during an on-line data taking of 7 days. A total of 26h 26 min was devoted to the study of $^{32}$Ar.
The average detection rate of $^{32}$Ar was about 100 pps.

\section{Experimental results}

In the following, we will first discuss the $\gamma$-ray spectra with $\gamma$ rays observed in the $\beta$-decay daughter nucleus $^{32}$Cl, in the
$\beta$p daughter $^{31}$S, and in the $\beta$ decay of $^{32}$Cl with $\gamma$ rays from $^{32}$S. In a second step, we will turn our attention to the emission
of protons with and without $\gamma$-ray coincidences in order to determine proton emission from excited states in $^{32}$Cl to the ground and first excited 
states of $^{31}$S. In contrast to Bhattacharya {\it et al.}, proton emission to higher excited states could not be observed in the present work.
We neglect the very weak $\beta$p (0.026\%) and $\beta\alpha$ (0.054\%) decay channels of $^{32}$Cl~\cite{firestone}, which might very slightly alter the total
proton emission branching ratio.

For the determination of average values, we use in the calculations one digit more than given in the tables. Therefore, a difference of 1 in the last digit
given may occur for calculations with the numbers in the table. If the different values to be averaged are more apart from each other than allowed by 
errors, we increase the error of the average value by the square root of the normalised $\chi^2_\nu$ as prescribed by the Particle Data 
Group~\cite{pdg18}.

\begin{figure}[hht]
\begin{centering}
\includegraphics[scale=0.4,angle=0]{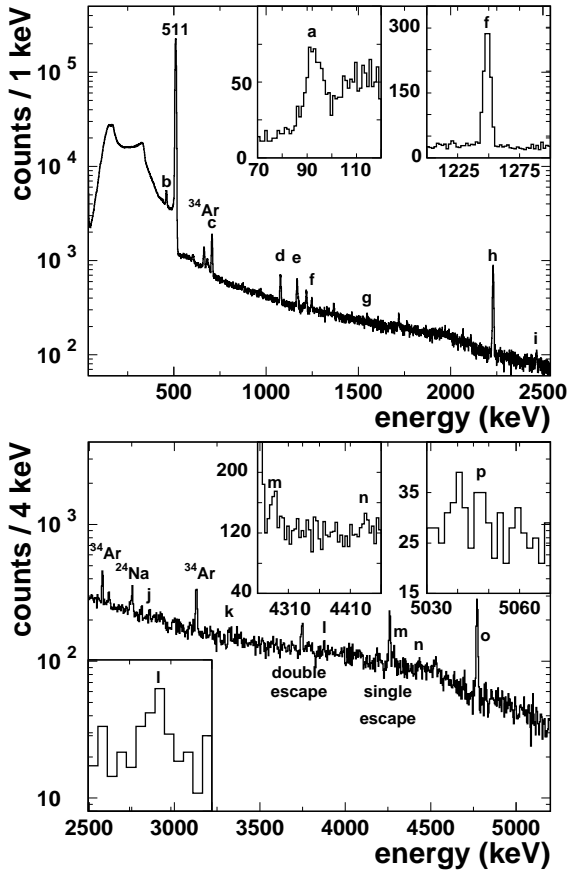}
\par\end{centering}
\caption{{\small{Gamma-ray spectrum as determined in the present work. Peaks from $a$ to $e$ as well as peaks $j$, $l$, $p$ belong to decays of excited states
                 of $^{32}$Cl. Peak $f$ stems from the decay of the first excited state to the ground state in $^{31}$S, whereas all other peaks belong to 
                 the decay of excited levels of $^{32}$S. Background lines are identified by the decaying nuclei. The binning has been changed 
                 between the upper and lower parts of the figure and for some of the inserts.
}}}
\label{fig:gamma}         
\end{figure}

\subsection{Gamma-ray spectra}

The first step in the data analysis is the determination of the $\gamma$-ray energies and their branching ratios. Figure~\ref{fig:gamma} shows the spectrum obtained 
from the eleven germanium segments used in the analysis. The peaks labelled with letters belong to the decay of $^{32}$Ar. Labels $a-e, j, l, p$, are from levels in 
$^{32}$Cl, in the $\beta$p daughter $^{31}$S (peak $f$) or from the decay of the $\beta\gamma$ daughter nucleus $^{32}$Cl ($g-i,k,n-o$). 
Peak $a$ is not visible in the $\gamma$-ray singles spectrum. It becomes visible only in the $\gamma-\gamma$ coincidence spectrum. In addition, 
it is partially cut by the ADC thresholds for some of the crystals. Therefore, it is difficult to determine its exact energy and 
branching ratio. We refrain from doing this, as its branching ratio is the same as the one of the $\gamma$ ray at 1078.6~keV (see decay scheme in figure~\ref{fig:scheme}). 
Peak $f$ is visible in the singles spectrum, but much better evidenced in a proton-$\gamma$-ray coincidence spectrum, as shown in the inset of the upper part of figure~\ref{fig:gamma}. 
The other insets show expanded views of the singles spectrum to better visualise the $\gamma$ rays of interest.

Table~\ref{tab:cl32_gam} summarizes the $\gamma$-ray data from this work and from Bj{\"o}rnstad {\it et al.}~\cite{bjornstad85} for states in the $\beta$-decay daughter $^{32}$Cl 
populated by a Gamow-Teller $\beta$ decay and for the $\gamma$ ray in the $\beta$p daughter $^{31}$S. Reasonable agreement 
is obtained for all energies. As the $\gamma$ lines from the decay of $^{32}$Ar were not included in the energy calibration, we can build the averages from the present work 
and from Bj{\"o}rnstad {\it et al.} For the branching ratios, good agreement is also obtained, except for the $\gamma$ ray at 1249~keV. The value from our work is overestimated, 
because this $\gamma$ ray is always in coincidence with protons, for which the trigger probability of the data acquisition is significantly higher than in the case 
of $\beta$-delayed $\gamma$ rays triggered by the $\beta$ particles in the DSSSDs. We refrain therefore from averaging the two values.

The absolute branching ratios can be obtained from the complement of the total proton emission branching ratio from Bhattachar\-ya {\it et al.} of BR$_p(tot)$~= 35.58(22)\%. 
We assume here that the ground-state feeding is negligible (see below). The results of this procedure are given in the last row of table~\ref{tab:cl32_gam}. 

Three $\gamma$ rays were observed in the work of Bhattacharya {\it et al.} to de-excite the IAS of the ground state of $^{32}$Ar in $^{32}$Cl. These are peaks $j$, $l$, and $p$, where we 
admit that peak $p$ would have been difficult to identify without knowledge from the work of Bhattacharya {\it et al.} Nevertheless, the energies and branching ratios determined 
show good agreement with the results of Bhattacharya {\it et al.} The energies as well as the absolute branching ratios and their averages are presented in table~\ref{tab:ias_gam}
and will be discussed below.

Gamma rays were not only observed in the $\beta$p daughter nucleus $^{31}$S, but also in the decay of the $\beta\gamma$ daughter nucleus $^{32}$Cl. These are given in table~\ref{tab:s32_gam}.
To obtain their absolute branching ratios, we normalised them with the total $\beta\gamma$ branching ratio calculated from the total proton branching ratio as given by 
Bhattacharya {\it et al.}~\cite{bhattacharya08}. In the table, we also present a comparison to data from the literature~\cite{nndcA=32}. Good agreement is obtained indicative of 
the fact that the extrapolated $\gamma$-ray detection efficiencies work well up to energies as high as 5~MeV.

\addtolength{\tabcolsep}{0pt}
\begin{table*}[htt]
\caption{{\small{}\label{tab:cl32_gam}}
         Energies as well as relative and absolute branching ratios for $\gamma$ rays from the decay of $^{32}$Ar. Results from the present work are compared to
         results from Bj{\"o}rnstad {\it et al.}~\cite{bjornstad85}.}
\begin{center}
\footnotesize{
\begin{tabular}{cccccccc}
\hline 
\hline 
\rule{0pt}{1.3em}
peak&\multicolumn{3}{c}{$\gamma$-ray energies (keV)} &\multicolumn{3}{c}{relative branching ratios (\%)}&  average absolute   \\
   &   this work&    Bj{\"o}rnstad \cite{bjornstad85} & average &
       this work&    Bj{\"o}rnstad \cite{bjornstad85} & average &  branching ratio (\%)\\
[0.5em]\hline 
\rule{0pt}{1.3em}
a~~&      -~~~~   &      89.9(1) &      89.9(1) &      38.3(32) &      36.6(18) &      37.0(16) &      13.8(7) \\
 b &     460.7(4) &     461.1(1) &     461.1(1) &     100.0(69) &     100.0(49) &     100.0(40) &      37.2(18) \\
 c &     707.1(4) &     707.4(2) &     707.3(2) &     100.0(69) &     100.0(49) &     100.0(40) &      37.2(18) \\
 d &    1078.1(4) &    1078.7(2) &    1078.6(2) &      38.3(32) &      36.6(18) &      37.0(16) &      13.8(7) \\
 e &    1168.3(5) &    1168.5(2) &    1168.5(2) &      31.8(38) &      36.9(18) &      35.9(20) &      13.4(8) \\
 f &    1248.5(5) &    1248.4(3) &    1248.4(3)\footnotemark &      17.1(17)\footnotemark &       5.8(6) &       5.8(6) &       2.2(2) \\
[0.5em]\hline 
\hline 
\end{tabular}
}
\end{center}
\footnotetext{1}{\hspace*{1cm}$^1$ In \cite{nndcA=31}, this energy is given as 1248.87(9)~keV. We use in the following text 1249~keV}\\
\footnotetext{2}{\hspace*{1cm}$^2$ $\beta$p triggered $\gamma$ ray, not used in averaging procedure due to detection efficiency difference, see text}
\end{table*}

\begin{table*}[hht]
\caption{{\small{}\label{tab:ias_gam}
         The table presents $\gamma$ rays de-exciting the IAS in $^{32}$Cl. The energies and branching ratios are averaged with similar values from the work of 
         Bhattacharya {\it et al.}~\cite{bhattacharya08}.}}
\begin{center}
\footnotesize{
\begin{tabular}{ccccccc}
\hline 
\hline 
\rule{0pt}{1.3em}
peak&\multicolumn{3}{c}{$\gamma$-ray energies (keV)} &\multicolumn{3}{c}{absolute branching ratios (\%)}  \\
   &   this work &Bhattacharya \cite{bhattacharya08}& average &
       this work &Bhattacharya \cite{bhattacharya08}& average\\
[0.5em]\hline 
\rule{0pt}{1.3em}
j~~&   2838.7(34) &    2836.0(10) &    2836.2(10) &  0.50(37) &  0.24(3) &  0.24( 3) \\
l &    3877.7(42) &    3877.5( 3) &    3877.5( 3) &  1.03(22) &  1.58(8) &  1.52(18) \\
p &    5047.5(50) &    5046.3( 4) &    5046.3( 4) &  0.22(14) &  0.10(2) &  0.10( 2) \\
[0.5em]\hline 
\hline 
\end{tabular}
}
\end{center}
\end{table*}

\addtolength{\tabcolsep}{8pt}

\begin{table*}[htt]
\caption{{\small{}\label{tab:s32_gam}}
         Gamma rays de-exciting levels in the $\beta$-decay grand daughter nucleus $^{32}$S. The branching ratios are compared with data 
         from the literature~\cite{nndcA=32}. The $\gamma$-ray energies have been used for the energy calibration.}
\begin{center}
\footnotesize{
\begin{tabular}{ccccc}
\hline 
\hline 
\rule{0pt}{1.3em}
peak& $\gamma$-ray energies (keV) &\multicolumn{3}{c}{absolute branching ratios (\%)}  \\
   &    & 
       this work &  \cite{nndcA=32}& average\\
[0.5em]\hline 
\rule{0pt}{1.3em}
g~~&     1548.0(20) &   2.4(4) &   2.7(5)  &   2.5(3)  \\
 h &     2230.5( 2) &  81.5(44)&  70.0(30) &  73.6(54) \\
 i &     2463.8(10) &   3.0(7) &   3.0(3)  &   3.0(3)  \\
 k &     3317.5(15) &   1.3(4) &   1.9(3)  &   1.7(3)  \\
 m &     4281.5(15) &   1.3(5) &   2.0(1)  &   2.0(1)  \\
 n &     4433.0(20) &   0.9(6) &   0.6(2)  &   0.6(2)  \\
 o &     4770.0(15) &  14.3(17)&  15.5(15) &  15.0(11) \\
[0.5em]\hline 
\hline 
\end{tabular}
}
\end{center}
\end{table*}

\subsection{Proton spectra}

\begin{figure}[hht]
\begin{centering}
\includegraphics[scale=0.3,angle=0]{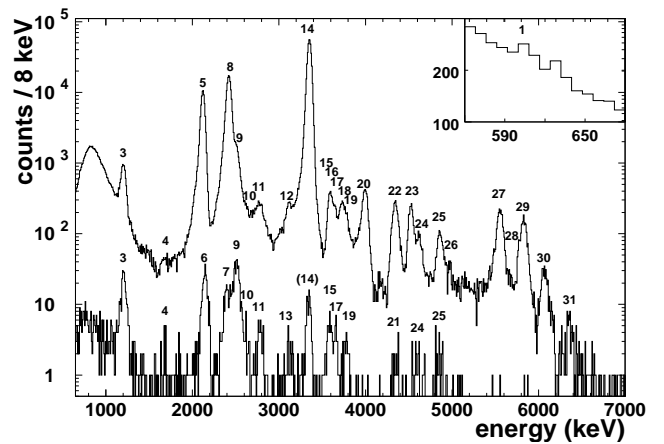}
\par\end{centering}
\caption{{\small{Proton energy spectrum for proton singles (higher-statistics spectrum) and in coincidence with a $\gamma$ ray from the first excited state of
                 $^{31}$S to its ground state. All peaks are labelled with numbers from 1 to 31 (see table~\ref{tab:protons}) used consistently throughout the present paper. 
                 Peak 2 is not observed in the present work nor in \cite{bjornstad85} (see \cite{bhattacharya08}).
}}}
\label{fig:protons}         
\end{figure}

Figure \ref{fig:protons} shows the singles proton spectrum from DSSSDs 1, 3 and 6. A spectrum was also created with the same proton spectrum, however, in coincidence
with a $\gamma$ ray at 1249~keV, the $\gamma$ ray de-exciting the first excited state of $^{31}$S to its ground state. The result is also shown in figure~\ref{fig:protons}. 
A total of 30 proton lines were observed in the present work. A very weak peak at 912~keV observed by Bhattacharya {\it et al.}~\cite{bhattacharya08}
was not seen in the present work nor in the work of Bj{\"o}rnstad {\it et al.}~\cite{bjornstad85}.

All proton lines were fit by a Gaussian and a low-energy tail to take into account the possibility that the full charge was not always collected in the silicon detectors or
that interference effects create a low-energy tail.
One single Gaussian plus a tail was sufficient for all proton groups to get a good description of the experimental peak, except for the line at 3353 ~keV, which is due to
the emission of a proton from the IAS, the strongest peak in the spectrum of figure~\ref{fig:protons}. In this case (peak 14), a second Gaussian was added.
The integral of the Gaussians yielded the number of counts in the different peaks and allowed the determination of relative branching ratios. In table~\ref{tab:protons}, 
we present the energy values of these proton lines and their relative branching ratios normalised to 1000 for the strongest proton line from the IAS.
The data from the present work are compared to the data available from the work of Bj{\"o}rnstad {\it et al.}~\cite{bjornstad85} and Bhattacharya 
{\it et al.}~\cite{bhattacharya08}, this latter work giving only proton lines with energies below about 4~MeV if they feed the first excited state in $^{31}$S (except for peak 1),
and all proton lines above 4~MeV.

\addtolength{\tabcolsep}{-11.5pt}

\begin{table*}[htt]
\caption{{\small{}\label{tab:protons}}
         The table gives the energies determined for the different proton groups from the present experiment, from Bhattacharya {\it et al.}~\cite{bhattacharya08}, 
         and from Bj{\"o}rnstad {\it et al.}~\cite{bjornstad85} and their averages. Similar data are presented for the branching ratios normalised to the most prominent peak at 3353~keV.
         The absolute proton intensities are obtained by means of the total proton emission branching ratio as determined by Bhattacharya {\it et al.} The peak numbers in the
         first column correspond to the peak labels in figure~\ref{fig:protons}. Proton groups 6 and 14 depopulate the IAS. Asterisks indicate peaks for a decay to the first excited state of $^{31}$S.
         There are some differences for the attribution to the decay to the ground or excited state in $^{31}$S between our work and the work of Bj{\"o}rnstad {\it et al.}.  They are indicated by
         a superscript, if these authors attributed the decay to the ground ("g") or excited ("e") state in constrast to us. All energies are in keV, all branching ratios in \%. 
         For details see text.
         }
\begin{center}
\scriptsize{
\begin{tabular}{lcccccccccc}
\hline 
\hline 
\rule{0pt}{1.3em}
   & \multicolumn{4}{c}{proton laboratory energies} & \multicolumn{4}{c}{relative branching ratios} & & \\
peak&        this   &   Bhattacharya  &   Bj{\"o}rnstad    &    average     &          this   &    Bhattacharya  &   Bj{\"o}rnstad     &     average     &     CM         &    absolute    \\
   &       work     & et al.~\cite{bhattacharya08} &  et al.~\cite{bjornstad85}&                &        work     &  et al.~\cite{bhattacharya08} &  et al.~\cite{bjornstad85} & 
                   &   energies     &     BR         \\
[0.5em]\hline 
\rule{0pt}{1.3em}
1      &     606.8( 60) &     610.0(100) &     607.0(100) &     607.5( 46) &       32.3(165)&       18.8( 4) &     17.0(31) &     18.76( 39) &     627.3( 46) &     0.385( 8) \\
~2$^*$ &          -     &     912.0( 50) &          -     &     912.0( 50) &          -     &      0.70(40) &          -     &      0.70( 40) &     941.6( 50) &     0.014( 8) \\
~3$^*$ &    1205.5( 41) &    1218.0( 50) &    1214.0(100) &    1210.8( 42) &      14.68(78) &       19.0(22) &     17.9(18)   &     15.57(112) &    1250.1( 42) &     0.319(23) \\
~4$^*$ &    1676.6(123) &          -     &          -     &    1676.6(123) &       1.43(65) &          -     &          -     &      1.43( 65) &    1731.1(123) &     0.029(13) \\
~5     &    2121.3( 40) &          -     &    2124.0(100) &    2121.7( 37) &      176.6(30) &          -     &    174.8(204)\footnotemark &    176.6( 30) &    2190.6( 37) &     3.62(7) \\
~6$^*$ &    2146.0( 52) &    2145.0( 50) &          -     &    2145.5( 36) &       18.7(40  &     12.80(40) &     12.9(18)\addtocounter{footnote}{-1}\footnotemark &     12.86(  41) &    2215.2( 36) &     0.264( 9) \\
~7$^*$ &    2395.9( 41) &    2394.0( 50) &          -     &    2395.1( 32) &        9.7(47) &      5.6(11) &          -     &      5.8(11) &    2472.9( 32) &     0.119(22) \\
~8     &    2422.9( 40) &          -     &    2423.0(100) &    2422.9( 37) &      352.8(47) &          -     &    358.1(393)&    352.8(47) &    2501.6( 37) &     7.24(11) \\
~9$^*$ &    2510.0( 40) &    2515.0( 50) &    2511.0(100)$^g$ &    2511.9( 30) &       36.2(12) &     29.3(11) &     39.3(87) &     32.6(25) &    2593.5( 30) &     0.668(51) \\
10$^*$ &    2615.7( 86) &          -     &          -     &    2615.7( 86) &       2.47(63) &          -     &          -     &      2.47( 63) &    2700.6( 86) &     0.051(13) \\
11$^*$ &    2778.5( 65) &    2870.0( 50)\footnotemark &    2768.0(100)$^g$ &     2775.4(54) &      3.83(50) &     30.0(100)\addtocounter{footnote}{-1}\footnotemark &      5.68( 87) &      4.29(  80) &    2865.6(54) &     0.088(16) \\
12     &    3117.4( 47) &          -     &    3113.0(100) &    3116.6( 43) &       1.32(13) &          -     &      1.67( 39)\addtocounter{footnote}{-2}\footnotemark &      1.36( 12) &    3217.8( 43) &     0.028( 2) \\
13$^*$ &    3117.0(105) &          -     &          -     &    3117.0(105) &       0.76(41) &          -     &      0.95( 27)\addtocounter{footnote}{-1}\footnotemark &      0.90( 22) &    3218.3(105) &     0.018( 5) \\
14     &    3352.7( 40) &    3353.0( 50) &    3353.5( 30) &    3353.2( 22) &     1000.0(41) &     1000.0(59)&   1000.0(699)  &   1000.0( 34) &    3462.1( 22) &    20.51(17) \\
15$^*$ &    3584.4(239) &    3581.0( 50) &    3592.0(100) &    3583.2( 44) &        3.5(19) &      2.40(40) &      3.48( 68)\addtocounter{footnote}{-1}\footnotemark &      2.70( 34) &    3699.6( 44) &     0.055( 7) \\
16     &    3605.2( 82) &          -     &          -     &    3605.2( 82) &       3.56(45) &          -     &      4.82( 69)\addtocounter{footnote}{-1}\footnotemark &      3.94( 58) &    3722.3( 82) &     0.081(12) \\
17$^*$ &    3651.8( 68) &    3649.0( 50) &    3643.0(100) &    3649.0( 37) &       3.21(37) &      3.20(30) &      3.93( 87) &      3.25( 23) &    3767.5( 37) &     0.067( 5) \\
18     &    3725.9( 48) &          -     &    3732.0(100) &    3727.1( 44) &       4.03(17) &          -     &      6.29( 83)\addtocounter{footnote}{-1}\footnotemark &      4.13( 46) &    3848.1( 44) &     0.085( 9) \\
19$^*$ &    3778.2( 63) &    3785.0( 50) &          -     &    3782.4( 39) &       2.01(33) &      5.20(50) &      3.32( 80)\addtocounter{footnote}{-1}\footnotemark &      3.00( 98) &    3905.2( 39) &     0.062(20) \\
20     &    3997.8( 50) &    3984.0( 50) &    3994.0(100) &    3991.2( 46) &       9.04(21) &     11.0(10) &     11.79( 87) &      9.26( 50) &    4120.8( 46) &     0.190(10) \\
21$^*$ &    4347.5( 96) &    4340.0( 50) &    4341.0(100)$^g$& 4341.5( 41) &       1.28(43) &      2.3(16)\addtocounter{footnote}{-1}\footnotemark &      1.44( 53)\addtocounter{footnote}{-1}\footnotemark &      1.38(  33) &    4482.5( 41) &     0.028( 7) \\
22     &    4344.9( 41) &    4340.0( 50) &    4341.0(100)$^g$& 4342.7( 30) &       4.94(18) &      4.8(14)\addtocounter{footnote}{-1}\footnotemark &      5.55( 91)\addtocounter{footnote}{-1}\footnotemark &      4.96(  18) &    4483.8( 30) &     0.102( 4) \\
23     &    4524.6( 41) &    4529.0( 50) &    4521.0(100)$^e$& 4525.9( 30) &       4.45(13) &      5.40(50) &  5.24( 44) &      4.57( 21) &    4672.8( 30) &     0.094( 4) \\
24$^*$ &    4622.0( 46) &    4630.0( 50) &    4621.0(100) &    4625.2( 32) &        1.5( 7) &      1.60(50) &      1.75( 44) &      1.65( 30) &    4775.4( 32) &     0.034( 6) \\
25$^*$ &    4862.5(180) &    4869.0( 50) &    4858.0(100) &    4866.5( 43) &        2.0(14) &      2.60(30) &      2.62( 44) &      2.59( 24) &    5024.6( 43) &     0.053( 5) \\
26     &    4966.7( 79) &    4997.0(100) &    4975.0(100) &    4977.5( 89) &       0.38(17) &      1.00(20) &      0.61( 17) &      0.63( 17) &    5139.1( 89) &     0.013( 4) \\
27     &    5556.1( 41) &    5567.0( 50) &    5552.0(100) &    5559.8( 40) &       5.39(13) &      7.60(80) &      5.68( 44) &      5.47( 25) &    5740.3( 40) &     0.112( 5) \\
28     &    5673.4(124) &    5699.0(100) &    5675.0(100) &    5683.7( 85) &       0.20( 5) &      1.80(80) &      0.48( 31) &      0.22(  8) &    5868.3( 85) &     0.005( 2) \\
29     &    5815.3( 44) &    5833.0( 50) &    5817.0(100) &    5822.5( 60) &       4.00(21) &      5.40(50) &      4.37( 44) &      4.23( 32) &    6011.6( 60) &     0.087( 7) \\
30     &    6066.2( 52) &    6097.0(100) &    6060.0(100) &    6070.5( 88) &       1.15(11) &      1.10(20) &      0.92( 17) &      1.09(  8) &    6267.7( 88) &     0.022( 2) \\
31     &    6358.6( 73) &    6396.0(100) &    6347.0(100) &    6365.2(133) &       0.56( 8) &      0.60(20) &      0.61( 17) &      0.57(  7) &    6572.0(133) &     0.012( 1) \\
[0.5em]\hline 
\hline 
\end{tabular}
}
\end{center}
\footnotetext{2}{\hspace*{0cm}$^3$ only one peak observed, sharing between ground and first excited states recalculated according to branching ratios from other experiments, see text}\\
\footnotetext{2}{\hspace*{0cm}$^4$ not used in the averaging procedure, see text}
\end{table*}

In the following, we quickly discuss all proton lines from table~\ref{tab:protons} individually. For this discussion, we will use the average proton energies.
The numbers correspond to the peak numbers used in figure~\ref{fig:protons} and in table~\ref{tab:protons}.

\begin{itemize}
\item
1: In the present work, we can see the 608 keV peak only under the condition that we require a signal in two different strips (one being from the proton, one from the 
associated $\beta$ particle) and keep only the higher energy. This condition drastically reduces the $\beta$ background at low energies and makes this peak visible. 
For a correct normalisation for this peak, the 3353~keV line was also analysed under the same condition. It is a proton to the ground state of $^{31}$S.
\item
2: The 912 keV line is not identified in our data. Only Bhattacharya {\it et al.} have a tiny proton branch at this energy to the first excited state.
\item
3: The 1211~keV line is seen in the three experiments as a proton emission to the first excited state in $^{31}$S with comparable intensities.
\item
4: The authors from previous work did not identify a proton line at 1677 keV, but we have clear evidence for such a line in the singles and in the coincidence spectra. 
Bj{\"o}rnstad {\it et al.} have also a small peak at this energy, however, they do not mention it. We attribute it to the decay to the first excited state of $^{31}$S.
\item
5-6: Bj{\"o}rnstad {\it et al.} identify a single peak at 2124~keV. However, in our data we have clear evidence for a ground-state decay at 2122~keV and a decay to the first excited state
at 2146~keV. This is in agreement with the work of Bhattacharya {\it et al.} who observe also a peak to the first excited state at 2146~keV. The branching ratio to the ground
state and to the first excited state as determined in the present work add up to the branching ratio given by Bj{\"o}rnstad {\it et al.} So we determine from the average of our work and the one of
Bhattacharya {\it et al.} the relative branching ratios to the ground and first excited states and use these value to distribute the branching ratio found by Bj{\"o}rnstad {\it et al.}
to the ground and first excited states for these authors.
\item
7: Bj{\"o}rnstad {\it et al.} have no peak at 2395 keV. In the present work and in Bhattacharya {\it et al.}, this peak is visible as a decay to the first excited state, although with almost 
a factor of 2 difference in  branching ratio.
\item
8: The peak at 2423~keV is observed in the present work and by Bj{\"o}rnstad {\it et al.} as a decay to the ground state in $^{31}$S.
\item
9: The 2512~keV line is observed in the three experiments with comparable intensity. As Bhattacharya {\it et al.}, we see it as a decay to the first excited state.
Bj\"ornstad {\it et al.} identify it as a decay to the ground state. Due to the clear evidence from the present work and from Bhattacharya {\it et al.}, we attribute it to
the decay to the first excited state.
\item
10: No 2616 keV peak is observed in the work of Bhattacharya {\it et al.} and of Bj{\"o}rnstad {\it et al.}, but there is a visible shoulder in the present data and in those of Bj{\"o}rnstad {\it et al.},
which we identify as a proton emission to the first excited state. The observation of this peak may be questionned due to the fact that there is no corresponding decay to the
ground state of $^{31}$S. This needs to be investigated in a new experiment.
\item
11-13: At 2775 and 3117 keV, Bj{\"o}rnstad {\it et al.} have two peaks as in the present work, but these authors do not give them in coincidence with the $\gamma$ ray at 1249~keV,
which could be due to the small branching ratio and their small $\gamma$-ray detection efficiency. We clearly see this coincidence and attribute these two peaks to decays to the first excited state.
However, in our data, the 3117 keV peak is also in the ground-state spectrum with a branching ratio of 1.32\% (peak 12).
Bhattacharya {\it et al.} see in this region a broad structure at 2870 keV with a much higher relative intensity (30(10)\%) as compared to 
5.9(7)\% for the present work and 8.3(10)\% for Bj{\"o}rnstad {\it et al.} 
The data of Bhattacharya {\it et al.} for the 2870~keV peak were not used for the averaging of the proton energies nor for the branching ratios. Like for peaks 5 and 6, we distribute 
the branching ratio of Bj{\"o}rnstad {\it et al.} for the 3117~keV line to the ground-state and excited-state decay.
\item
14: The 3353 keV peak is the most prominent peak for all three data sets. It is therefore taken for the normalisation.  
As will be discussed below, a small contribution in coincidence with the $\gamma$ ray de-exciting the first excited state in $^{31}$S cannot be completely
excluded.
\item
15-16: Around 3600~keV, we see a somewhat broader peak with one part decaying to the ground state and one to the first excited state of $^{31}$S. 
We attribute the 3583~keV line to the decay to the first excited state and the 3605~keV line to the decay to the ground state.
Bhattacharya {\it et al.} see the decay to the first excited state, whereas Bj\"ornstad {\it et al.} see one peak "predominantly" decaying to the first excited state, 
indicative that in their data there might be also decay strength to the ground state. We therefore distribute their decay branching ratio to both decays
according to the relative branching ratios from our data and from Bhattacharya {\it et al.}
\item
17: Nice agreement is obtained for all three data sets for the peak at 3649 keV as a decay to the first excited state.
\item
18-19: The peak at 3732 keV in Bj{\"o}rnstad {\it et al.} has at least two components. We find a ground-state peak at 3727 keV and an excited-state peak at 3782 keV,
in agreement with Bhattacharya {\it et al.} If we sum the two peaks observed here for decays to the ground and first excited states, we come close to the integral 
of Bj{\"o}rnstad {\it et al.}. We distribute their decay branching ratio to both decays according to the relative branching ratios from our data and from Bhattacharya {\it et al.}
\item
20: The 3991 keV peak is observed in all three experiments with comparable intensities as a decay to the ground state.
\item
21-22: For the peak at about 4340 keV, we clearly see it for the ground state decay (22) and in the coincidence spectrum (21). So we share its activity between the ground state 
and the first excited state also for Bhattacharya {\it et al.} and Bj{\"o}rnstad {\it et al.}
\item
23: The 4526 keV peak is seen in the three experiments, but Bj{\"o}rnstad {\it et al.} claim to observe it predominantly in coincidence with the 1249~keV $\gamma$ ray, whereas 
Bhattacharya {\it et al.} see it mainly as a ground-state decay. We can only observe a ground-state branch at this energy. 
We attribute the strong ground-state decay observed by Bhattacharya {\it et al.} (0.54(5)\% with respect to the IAS decay) to this peak 23. We do the same for the
peak of Bj\"ornstad {\it et al.} (0.12(1)\% absolute branching ratio). We neglect the small possible excited-state contribution ($<$0.03\% with respect to the IAS peak 14)
from Bhattacharya {\it et al.}
\item
24-25: A proton group at 4625 keV is seen in the three experiments with comparable intensities as a decay to the
first excited state. The 4867 keV peak decays also to the first excited state.
\item
26-31: The proton groups at 4978, 5560, 5684, 5823, 6071, and 6365 keV are identified as decays to the ground state in $^{31}$S in all three experiments.
\end{itemize}

The last three peaks were evaluated with DSSSD 5. We found that the branching ratio determined with detectors 1, 3, and 6 was the same as with detector 5 within the error bars for proton
group 29, but factors of 2 and 6 lower for the high-energy proton groups 30 and 31, respectively. 

We note that the proton peak energies of Bhattacharya {\it et al.} are systematically higher than those given for our data and those from Bj\"ornstad {\it et al.}
above about 4500~keV. Schardt and Riisager~\cite{schardt93} mention that the energies of Bj{\"o}rnstad {\it et al.} should be corrected, because these authors used an $\alpha$-particle
energy calibration at high energies. However, the correction proposed moves the proton energies of Bj{\"o}rnstad {\it et al.} even above the values of Bhattacharya {\it et al.} 
and far away from our proton energies. For the present paper, we keep the energy values from Bj{\"o}rnstad {\it et al.}

The doublets that we observe at around 3130~keV, 3595~keV, and 4340~keV have been given particular attention. We first fit a single proton peak in the $\gamma$-coincidence
spectrum and in the ground-state decay spectrum, this latter being obtained by subtracting the spectrum for the decay to the first excited state with the correct normalisation
from the proton singles spectrum. This procedure was verified by fitting a doublet to the proton singles spectrum by keeping one of the two energies and the width of the peaks fixed.
It allowed us also to determine the relative intensities of the ground-state and excited-state decays and to observe that none of the two alone gives the intensity
of the singles-spectrum peak, when fitted with only one proton line.

The relative branching ratios were averaged (with the restrictions mentioned above) as given in table~\ref{tab:protons}. With the technique used in the present work, we cannot 
determine the number of implanted $^{32}$Ar in any precise way to transform the relative branching ratios into absolute proton emission branching ratios. In order to determine 
absolute branching ratios, we use again the total proton emission branching ratio of Bhattacharya {\it et al.} However, if we just add the integrals of the identified peaks
to determine the total number of protons detected in the present work, the procedure is not correct. Some activity is not identified in peaks (e.g. activity between peaks 26 and 27). 
We therefore determined the total number of protons by fitting an exponential to the low-energy $\beta$-particle contribution in the upper spectrum of figure~\ref{fig:protons} 
and subtracted this contribution before determining the total number of counts in the spectrum, which corresponds to the total number of protons detected. If we compare this number 
to the sum of the integrals of all identified proton groups, we find that a small correction factor of 1.033(14) is needed. With this correction and the total proton emission branching 
ratio from Bhattacharya {\it et al.} of BR$_p(tot)$~= 35.58(22)\%, we can determine the absolute branching ratios given in the table.

From the number of protons in the singles spectrum after the subtraction of the $\beta$-particle contribution at low energies and from the number of protons in the spectrum 
conditioned by a $\gamma$ ray at 1249~keV after subtraction of random coincidences (see below), we can determine the proton branching ratio to the first excited state 
of $^{31}$S to be 2.5(3)\%, which can be compared to the value obtained by Bhattacharya {\it et al.} of 2.3(4)\%. Bj{\"o}rnstad {\it et al.} found a value of 1.9(2)\%, whereas
we determine 2.1(2)\% with a slightly different absolute normalisation for the latter data as already discussed by Bhattacharya {\it et al.} (see table~\ref{tab:cl32_gam}).

\subsection{Proton-$\gamma$ coincidence data}

Up to now, we have treated the proton and the $\gamma$-ray data almost independently. Only to better evidence certain proton groups most likely being due to a decay to the
first excited state in $^{31}$S, we used proton-$\gamma$-ray coincidences to produce the lower-statistics spectrum shown in figure~\ref{fig:protons}. However, the fact that a peak 
shows up in this $\gamma$-ray coincidence spectrum does not necessarily mean that the protons are really in coincidence with the $\gamma$ ray at 1249~keV, although, 
as we will show, this is the case most of the time. We note also that in our experiment, due to the total number of decays observed and due to our efficiencies, the detection
limit for proton-$\gamma$ coincidences with one count corresponds to a branching ratio of 0.0015\%. However, for higher-intensity proton peaks the accidental coincidence probability is much higher.

\begin{figure}[hht]
\begin{centering}
\includegraphics[scale=0.39,angle=-90]{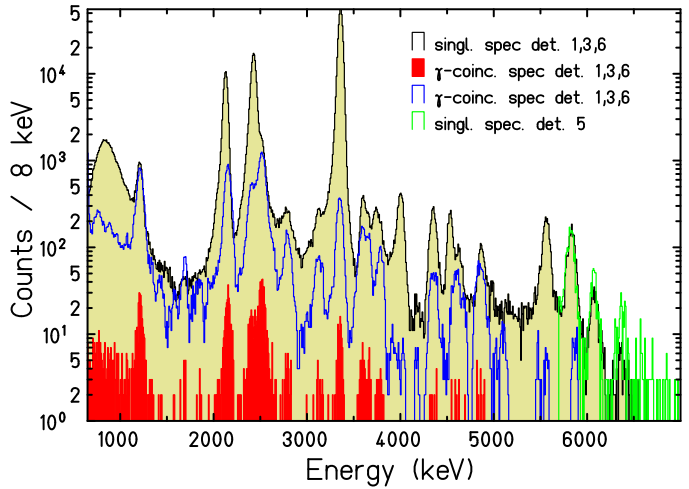}
\par\end{centering}
\caption{{\small{The figure shows the proton singles spectrum (black) overlaid with the proton spectrum in coincidence with a $\gamma$ ray at 1249~keV (red) and the same coincidence spectrum
                 multiplied with a factor of 35.1 to take into account the $\gamma$-ray detection efficiency (blue). The multiplied spectrum was folded with a Gaussian ($\sigma$~= 
                 2~keV) to smoothen the spectrum. These spectra were obtained with data from DSSSD 1, 3, and 6. The green spectrum plotted from 5700~keV onwards is a singles spectrum
                 generated from DSSSD~5 with a normalisation factor of 3 with respect to the other spectra for the different number of detectors.
}}}
\label{fig:p_gam}         
\end{figure}

In order to evidence proton-$\gamma$-ray coincidences, we overlay in figure~\ref{fig:p_gam} the proton singles spectrum from DSSSD 1, 3, and 6 as well as from DSSSD~5 with a
normalisation factor of 3 with the proton spectrum 
generated from the same detectors requiring a coincidence with a $\gamma$ ray between 1240~keV and 1254~keV (we do not subtract for the moment random coincidence, see below).
However, as the $\gamma$-ray detection efficiency decreases the number of counts observed in this coincidence spectrum, we multiplied the number of counts in each channel by a factor
of 35.1(19) determined as the average of the inverse of the $\gamma$-ray detection efficiency at 1249~keV and the ratios of the number of counts of the 1211~keV and the 2512~keV
proton peaks in singles and with the $\gamma$-ray coincidence. As just multiplying this spectrum with a factor of 35.1 would yield a spectrum difficult to "read", we 
"randomised" the multiplied spectrum by folding each event with a Gaussian with a sigma of 2~keV to smoothen the spectrum. The result is shown in figure~\ref{fig:p_gam} 
together with the proton singles spectra and the original $\gamma$-ray coincidence proton spectrum.

In these spectra, one clearly sees that for some proton groups one recovers almost the full proton singles intensity with the scaled coincidence spectrum (e.g. protons
at 1211~keV, 2512~keV, 4867~keV). However, there is also significant proton activity under the peak at 3353~keV, which is supposed to be the decay of the IAS in $^{32}$Cl to the
ground state of $^{31}$S. So this peak should not be in coincidence with $\gamma$ rays at 1249~keV.

\begin{figure}[hht]
\begin{centering}
\includegraphics[scale=0.3,angle=0]{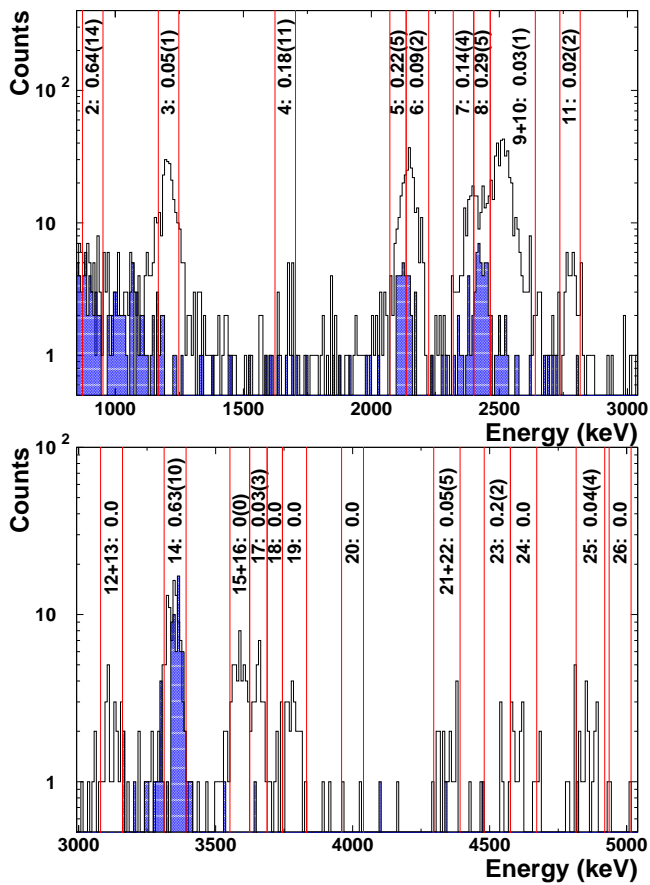}
\par\end{centering}
\caption{{\small{Proton spectra with a $\gamma$-ray coincidence at 1249~keV (black histogram) and with a coincidence left and right of this $\gamma$-ray energy 
                 (full blue spectrum). This latter spectrum contains only random coincidences. The numbers represent the peak label of figure~\ref{fig:protons}
                 and table~\ref{tab:protons} and ratios of the numbers of counts in the two spectra for a given interval. No counts are observed in the spectrum
                 with the random coincidences above 5000~keV.}}}
\label{fig:ratio_p_gam}         
\end{figure}

In order to determine the "real" coincidence counts and remove random coincidences, we show in figure~\ref{fig:ratio_p_gam} the same $\gamma$-ray coincidence spectrum
with the 1249~keV $\gamma$ ray and overlay a spectrum generated with a coincidence left and right of the 1249~keV peak. This latter spectrum should consist basically only of
random coincidences. In addition, we determine the ratios between the number of counts in both spectra for 21 proton-peak regions, which include 25 of our 31 proton groups. 
For the analysis of the ratios in figure~\ref{fig:ratio_p_gam}, we centered the regions, given by the vertical lines, at the energies of the proton groups 
(mainly for decays to the first excited state) with a width of $\pm$40~keV in the low-energy region and
$\pm$50~keV for higher proton energies (the last 4 regions). Peaks 9-10, 12-13, 15-16, and 21-22 were evaluated within a single region. For some peaks, the intervals are smaller than
$\pm$ 40~keV because two neighbouring peaks are too close in energy.

This evidences that we deal with 
almost the same number of counts e.g. for the region of the 3353~keV peak in both spectra. For regions with real coincidences (e.g. the peaks at 1211~keV, 2146~keV, 2775~keV, etc.), 
this ratio is small. However, often ground-state decays and excited-state proton groups overlay or are very close. A good ratio limit seems to be a value of 0.2 (see below), 
where real coincidence peaks have a smaller and random coincidence peaks have a larger value. From these ratios, we determine that peaks with numbers 3, 4, 6, 7, 9, 10, 
11, 13, 15, 17, 19, 21, 24, and 25 are proton emissions to the first excited state of $^{31}$S. For peak 2, the situation is more difficult to analyse due to the presence of a strong $\beta$ background. 
According to Bhattacharya {\it et al.}, it is a rather weak proton emission to the first excited state, which is not observed neither in our work nor by Bj{\"o}rnstad 
{\it et al.} We keep it as a proton emission to the first excited state of $^{31}$S.

\addtolength{\tabcolsep}{8pt}
\begin{table*}[htt]
\caption{{\small{}\label{tab:excit}}
         The first column gives the excitation energy in $^{32}$Cl followed by the absolute branching ratio, the B(GT)/B(F) values, and the proton group(s) de-populating these levels. 
         If two proton energies are given, decays to the ground and first excited states in $^{31}$S were observed. The numbers in brackets behind the center-of-mass proton energies 
         are the proton peak labels from table~\ref{tab:protons} and figure~\ref{fig:protons}. The first level indicates the bound level
         decaying by $\gamma$ decay. Its branching ratio is calculated from the total proton branching ratio and the $\gamma$ decay of the IAS. 
         The branching ratios do not sum up to 100\%, because some proton-emission strength could not be attributed to identifiable peaks.
         The excitation energy of the IAS (shown with bold letters) at 5046.1~keV is determined from the energies of its $\gamma$ decay.}
\begin{center}
\normalsize{
\begin{tabular}{ccccc}
\hline 
\hline 
\rule{0pt}{1.3em}
excitation  &     branching  &  B(GT)/ &  c.m. proton &  c.m. proton   \\
energy (keV)&     ratio (\%) &   B(F)  &  energy g.s. (keV)&energy exc. state (keV)\\
[0.5em]\hline 
\rule{0pt}{1.3em}
   1168.5( 2)  &     62.57( 29) & 0.4704( 99)  &        -        &     -           \\
   2208.4( 46) &     0.145( 31) & 0.0020(  4)  &     627.3  [ 1] &     -           \\
   3771.7( 30) &     3.635( 68) & 0.1387( 38)  &    2190.6  [ 5] &     941.6  [ 2] \\
   4081.6( 28) &     7.555(112) & 0.3644( 92)  &    2501.6  [ 8] &    1250.1  [ 3] \\
   4561.1(123) &     0.029( 13) & 0.0021(  9)  &        -        &    1731.1  [ 4] \\
   4799.0( 43) &     0.028(  2) & 0.0024(  2)  &    3217.8  [12] &     -           \\
\bf{5046.1( 3)}&  \bf{22.63(25)}&\bf{3.991(93)}&\bf{3462.1  [14]}& \bf{2215.2 [ 6]}\\
   5303.0( 30) &     0.200( 25) & 0.0276( 35)  &    3722.3  [16] &    2472.9  [ 7] \\
   5425.4( 27) &     0.753( 52) & 0.1169( 84)  &    3848.1  [18] &    2593.5  [ 9] \\
   5530.6( 86) &     0.051( 13) & 0.0087( 22)  &        -        &    2700.6  [10] \\
   5699.3( 35) &     0.278( 19) & 0.0569( 41)  &    4120.8  [20] &    2865.6  [11] \\
   6063.6( 45) &     0.120(  6) & 0.0365( 19)  &    4483.8  [22] &    3218.3  [13] \\
   6254.0( 31) &     0.094(  4) & 0.0355( 18)  &    4672.8  [23] &     -           \\
   6529.6( 44) &     0.055(  7) & 0.0293( 38)  &        -        &    3699.6  [15] \\
   6597.5( 38) &     0.067(  5) & 0.0385( 28)  &        -        &    3767.5  [17] \\
   6732.8( 56) &     0.074( 20) & 0.0514(141)  &    5139.1  [26] &    3905.2  [19] \\
   7317.0( 45) &     0.140(  8) & 0.2269(144)  &    5740.3  [27] &    4482.5  [21] \\
   7449.4( 85) &     0.005(  2) & 0.0090( 34)  &    5868.3  [28] &     -           \\
   7602.9( 73) &     0.121(  9) & 0.3147(244)  &    6011.6  [29] &    4778.1  [24] \\
   7853.5( 39) &     0.075(  5) & 0.3123(229)  &    6267.7  [30] &    5024.6  [25] \\
   8153.1(133) &     0.012(  1) & 0.0903(114)  &    6572.0  [31] &     -           \\
[0.5em]\hline 
\hline 
\end{tabular}
}
\end{center}
\end{table*}

Another finding that might deserve particular attention in a future high-resolution study is the fact that, for the 3353~keV peak number 14, the random coincidences are exclusively
located in the high-energy part of this peak. If one subtracts the random coincidences, a peak in the low-energy tail of this strong peak remains, which could be indicative
of a small proton peak with a real coincidence with the 1249~keV $\gamma$ ray. This might be important because it reduces slightly the Fermi strength discussed below.

The ratio of 0.2 used above might seem somewhat arbitrary, but it is meant to be an indication only for the ratio of counts in the different regions between the two spectra.
One can also subtract the spectrum with the random coincidence (condition left and right of the 1249~keV peak) from the coincidence spectrum (condition on the 1249~keV peak).
Unfortunately, the statistics becomes rather limited. However, if we fit this difference spectrum, which should contain only real coincidences, with energies fixed to the values
of table~\ref{tab:protons}, the numbers of counts for peaks without real coincidences (i.e. peaks 5 and 8) are compatible with zero and confirm thus the above conclusions.
As stated in the previous paragraph, counts remain in the low-energy part of peak 14.

\section{Results and discussion}

With the data analysed and presented in the previous sections, we can determine the probabilities of $\beta$-decay feeding of different states in $^{32}$Cl and
their decay to the ground or first excited state of $^{31}$S. From this information, we can build the decay scheme of $^{32}$Ar. 
We will then establish the Gamow-Teller strength distribution and compare it to shell-model predictions. Finally, we will analyse
the Fermi strength determined from the proton and $\gamma$ decay of the IAS.

In the following, we will assume that the decay branch from $^{32}$Ar to the ground state of $^{32}$Cl is negligibly small. This is justified, if mirror symmetry holds even
only rough\-ly, as the $log(ft)$ value for the mirror decay is as large as 8.21~\cite{nndcA=32} corresponding to a branching ratio for $^{32}$Ar decay to the ground state of 
$^{32}$Cl of 5$\times$10$^{-5}$. The different USD-type Hamiltonians (see below) give branching ratios ranging from 0.007\% (USD~\cite{brown88}) to 5.6\% (USDA~\cite{brown06}).

\subsection{Branching ratios to states in $^{32}$Cl and decay scheme of $^{32}$Ar}

In the preceding sections, we have been able to attribute all proton groups to decays from states in $^{32}$Cl to two states in $^{31}$S, the ground state and the 
first excited state at 1249~keV. No decay to other states of $^{31}$S could be evidenced. Proton energies having a center of mass difference
in agreement with a $\gamma$-ray energy of 1249~keV (we required a difference of less than 1.5 times the sum of the uncertainties of the proton energies) 
stem from the same level in $^{32}$Cl, but decay to the ground and first excited states of $^{31}$S. Table~\ref{tab:excit}
summarizes this information and gives the excitation energies of states populated in the decay of $^{32}$Ar (calculated with a proton separation energy of 
$S_p$~= 1581.1(5)~keV~\cite{ame2016}), the branching ratio to these levels and the center-of-mass energies of protons involved in the decay of these levels to the 
ground or first excited state in $^{31}$S.

If we compare the levels deduced in the present work with the adopted levels from the evaluation of states in $^{32}$Cl~\cite{nndcA=32}, we propose one additional level at 5530.6~keV,
whereas for four states from \cite{nndcA=32} we cannot identify the corresponding proton peaks (excitation energies 2611~keV, 4167~keV, 4439~keV, 5794~keV).
The first level comes from a "private communication". This level is observed in none of the $\beta$-decay experiments. However, it would correspond to a level clearly present in the mirror
nucleus $^{32}$P at 2740~keV. The second and third states come from the work of Bj\"ornstad {\it et al.} from two proton branches attributed to a ground-state decay. We clearly identify the two proton branches
as decays to the first excited state of $^{31}$S. The last level comes from the very broad 2870~keV proton line identified by Bhattacharya {\it et al.} only and discarded in our
analysis (see above). So from the four levels, we will keep only the 2611~keV line.

In addition to the unbound levels, which decay by proton emission, we give also the feeding of the bound 1$^+$ level at 1168.5 keV of $^{32}$Cl, which $\gamma$-decays 
to the ground state of $^{32}$Cl. This decay takes place via a $\gamma$-ray cascade that can be determined with the $\gamma$ rays observed and attributed to the decay 
of levels of $^{32}$Cl (see table~\ref{tab:cl32_gam}).

\begin{figure*}[hht]
\begin{centering}
\includegraphics[scale=0.6,angle=-90]{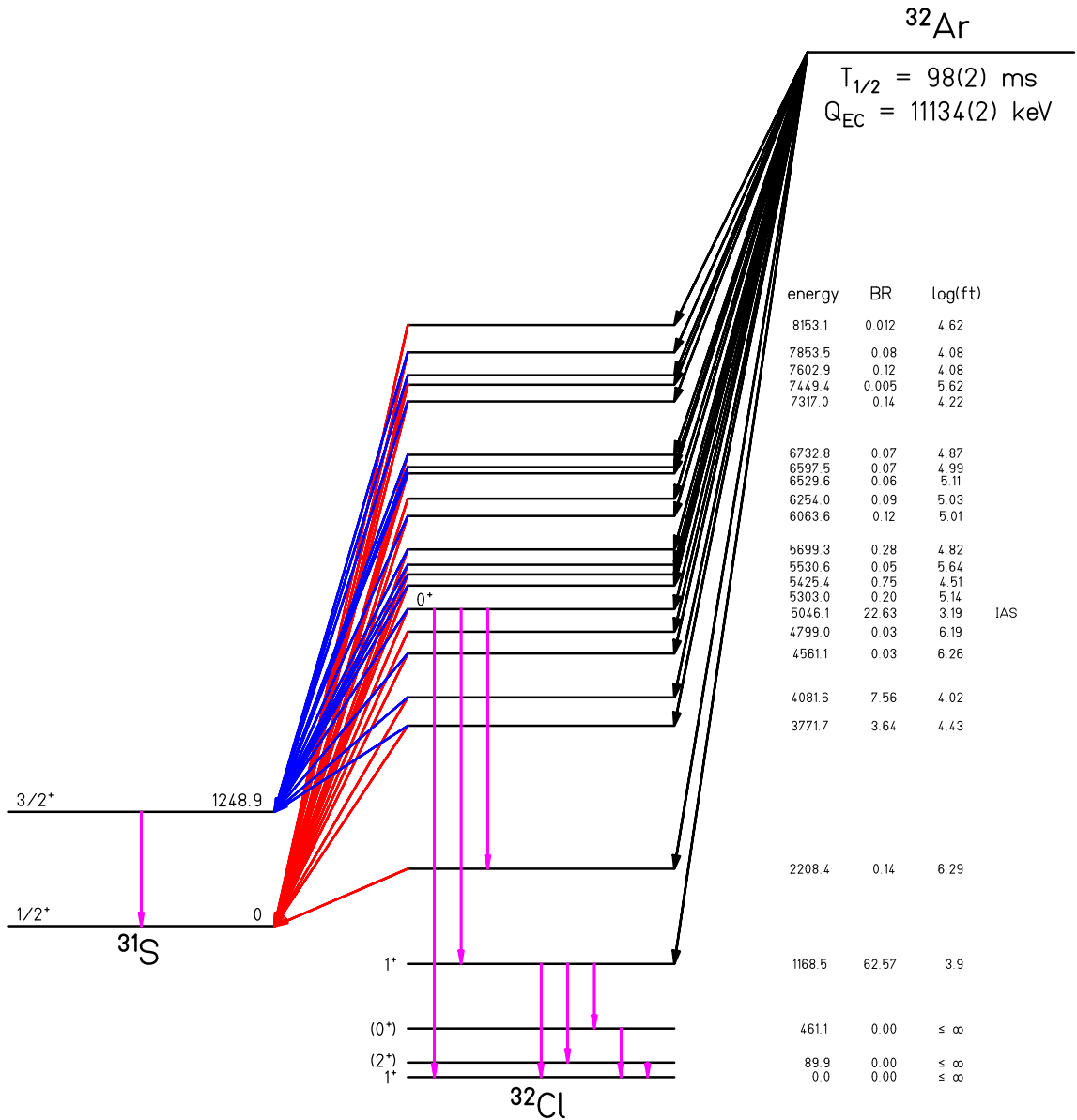}
\par\end{centering}
\caption{{\small{The decay scheme presented summarizes the results obtained in the present work averaged whenever possible with results from the work of Bj{\"o}rnstad {\it et al.}
                 \cite{bjornstad85} and of Bhattacharya {\it et al.}~\cite{bhattacharya08}. The proton separation energy used is $S_p$~= 1581.1(5)~keV~\cite{ame2016}. The characteristics 
                 for the decay of $^{32}$Ar are from the literature~\cite{nndcA=32}.}}}
\label{fig:scheme}         
\end{figure*}

All this information allows us finally to establish a rather complete decay scheme for $^{32}$Ar given in figure~\ref{fig:scheme}. Information from the
present paper is combined with the results from Bj{\"o}rnstad {\it et al.} and Bhattacharya {\it et al.} and data from the literature e.g. for the half-life and the $Q$ 
value~\cite{nndcA=32}. The branching ratios, the $\beta$-decay Q values and the half-life of $^{32}$Ar allowed us also to determine the $log(ft)$ values
for the different decay branches as given in the decay scheme.

\subsection{Gamow-Teller strength distribution}

The information gathered to establish the decay scheme can also be used to determine the Gamow-Teller strength distribution B(GT).
However, this procedure is only correct, if all decay strength can be included. As stated above, about 3.3\% of the total decay strength from unbound levels in $^{32}$Cl could not 
be attributed to a proton peak. This induces certainly only a small error in the B(GT) distribution, in particular at high excitation energy. 
We will determine this strength distribution in the standard way by using the identified peaks and their branching ratios as well as in a more precise way
by using directly the proton spectrum. 

For this latter purpose, we use the proton singles spectrum (from detectors 1, 3, and 6 up to 5.70~MeV, from detector 5 above), subtract the 
$\gamma$-coincident spectrum from the same detectors (multiplied with the factor of 35.1) to obtain the spectrum of protons to the ground state of $^{31}$S, take
the $\gamma$-coincident spectrum and shift it by 1249~keV, and add it back to the ground-state spectrum with the correct multiplication factor. 
This new spectrum can then be converted into a
spectrum of population of excited states in $^{32}$Cl. We next exclude the region of the IAS. The total number of counts in the spectrum obtained together 
with the total proton branching ratio allow us to give to each event and thus to each channel of this spectrum a branching ratio. The excitation energy, 
to which each channel corresponds, enables us to determine the $Q$ value for each channel and then the Gamow-Teller strength B(GT) as
$$
B(GT) = K / (G_A^2 * ft)
$$
with $K$ = 6143.6~s (see e.g.~\cite{hardy20}) and the axial-vector coupling constant $G_A$ = -1.2756~\cite{pdg20}. 
$f$ is the statistical rate function~\cite{towner15} depending on the $\beta$-decay $Q$ value and $t$ is the partial half-life defined as 
$$
t = T_{1/2} / BR
$$
with BR being the branching ratio attributed to each spectrum channel. The B(GT) value for the feeding of the bound state at an excitation energy of 1169~keV 
is calculated in a similar way. We used T$_{1/2}$~= 98(2)~ms~\cite{nndcA=32}.

\begin{figure}[hht]
\begin{centering}
\includegraphics[scale=0.46,angle=0]{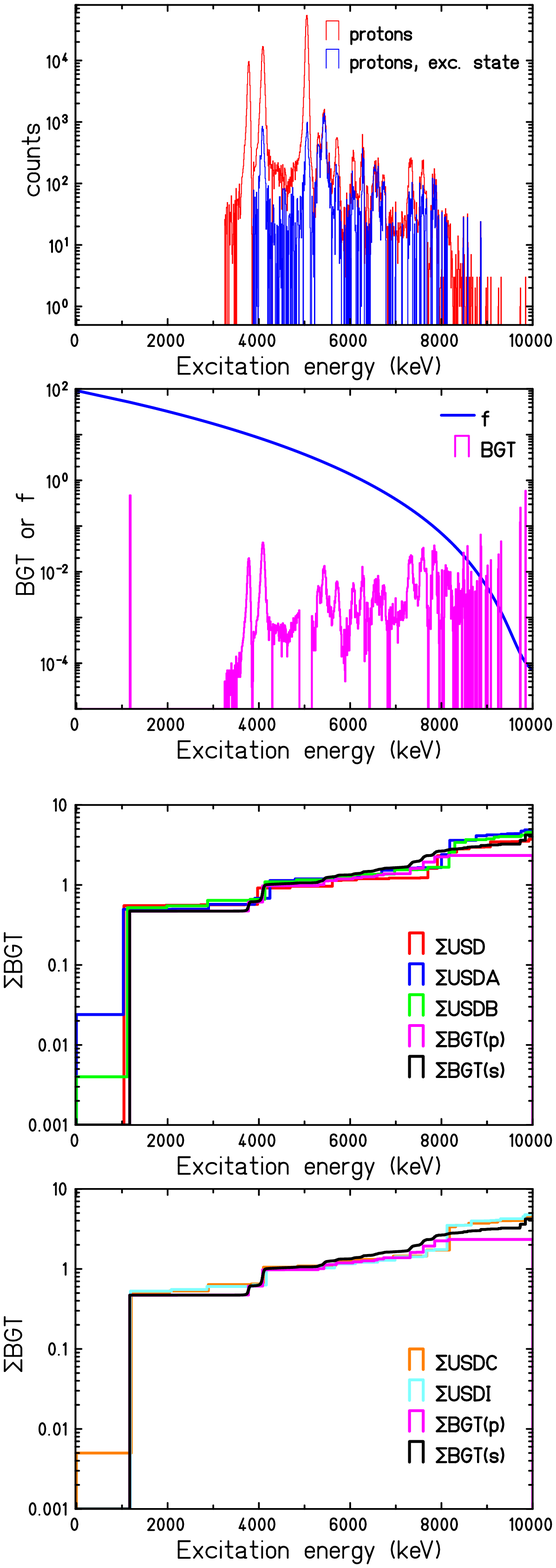}
\par\end{centering}
\caption{{\small{Upper part: the proton ground-state and excited-state spectra are converted to a spectrum as a function of the excitation energy in $^{32}$Cl.
                 The red spectrum shows the resulting excitation energy spectrum, whereas the blue spectrum is the contribution only from the first excited state in $^{31}$S.
                 Central part: the statistical rate function is shown together with the spectrum from the upper part divided by the statistical rate function.
                 We have also added here the decay to the bound state in $^{32}$Cl.
                 Lower part: the summed Gamow-Teller strength distribution obtained in the present work ($s$ for spectrum, $p$ for peaks) is compared to shell-model 
                 calculations with USD~\cite{brown88}, USDA, USDB~\cite{brown06}, and USDC, USDI~\cite{magilligan20} hamiltonians with a quenching factor of about 0.6. 
                 See text for details.
}}}
\label{fig:bgt}         
\end{figure}

The result of this procedure is shown in figure~\ref{fig:bgt} together with the B(GT) distribution determined with the identified proton peaks. One can see that starting from
about 6~MeV excitation energy, the B(GT) distribution obtained from the identified proton peaks stays below the B(GT) distribution from the full proton spectrum.

\subsection{Comparison with shell-model calculations}

We compare our results to calculations obtained in the {\it sd}-shell model space with protons and neutrons in the $(1s_{1/2},0d_{5/2},0d_{3/2})$ set of orbitals.
We will use the so-called ``universal" {\it sd} (USD) type Hamiltonians. Universal means that the same set of single-particle energies and two-body matrix elements
can be applied to all states whose wavefunctions are thought to be dominated by {\it sd}-shell model configurations. The two-body matrix elements
are allowed to have a smooth mass dependence. This is in contrast with the {\it ab-initio} in-medium similarity renormalization group (IMSRG) \cite{stroberg17,stroberg19}
method where the Hamiltonian is nucleus dependent. At present, the rms deviation between the experimental and theoretical excitation energies for IMSRG is about 500 keV.
Making this universal assumption, one can fine-tune the two-body matrix elements to minimize the deviation between experimental and theoretical energies and 
reduce the rms deviation down to about 150 keV.

We will use five USD-type Hamiltonians: (i) the original isospin-conserving USD Hamiltonian based on data for {\it sd}-shell nuclei with $N \geq Z$ available
up to 1983 \cite{brown88,wildenthal84}, (ii) the improved isospin-conserving USDA and USDB Hamiltonians based on data for nuclei with $N \geq Z$ available
up to about 2006 \cite{brown06}, and (iii) the most recent isospin-nonconserving USDC and USDI Hamiltonians based on data for all {\it sd}-shell nuclei \cite{magilligan20}.
The isospin-nonconserving part contains the Coulomb interaction and an isotensor strong interaction.

\subsubsection{Comparison of experiment and theory for $^{32}$P}

\begin{figure*}[hht]
\begin{centering}
\includegraphics[scale=0.48,angle=0]{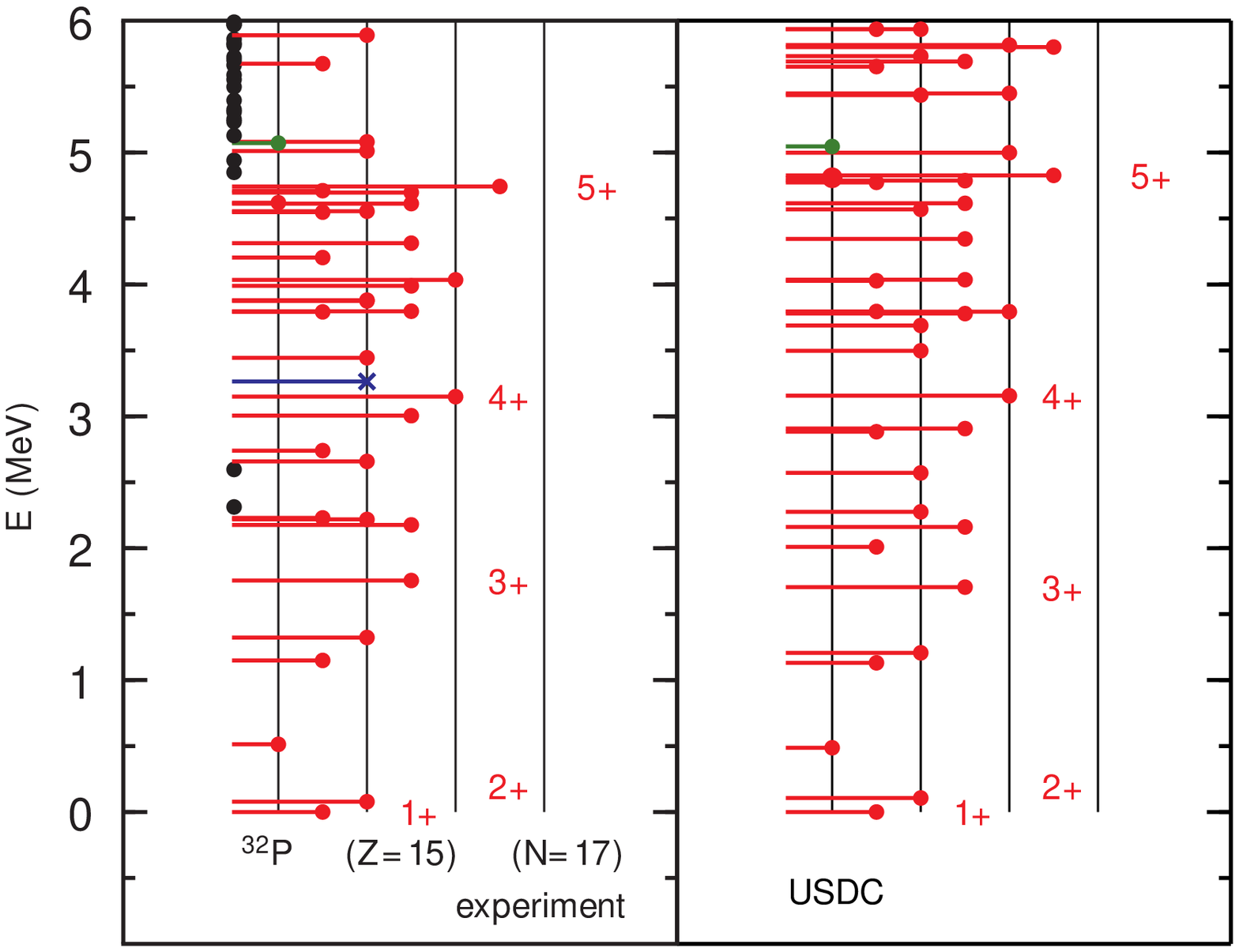} \hfill
\includegraphics[scale=0.48,angle=0]{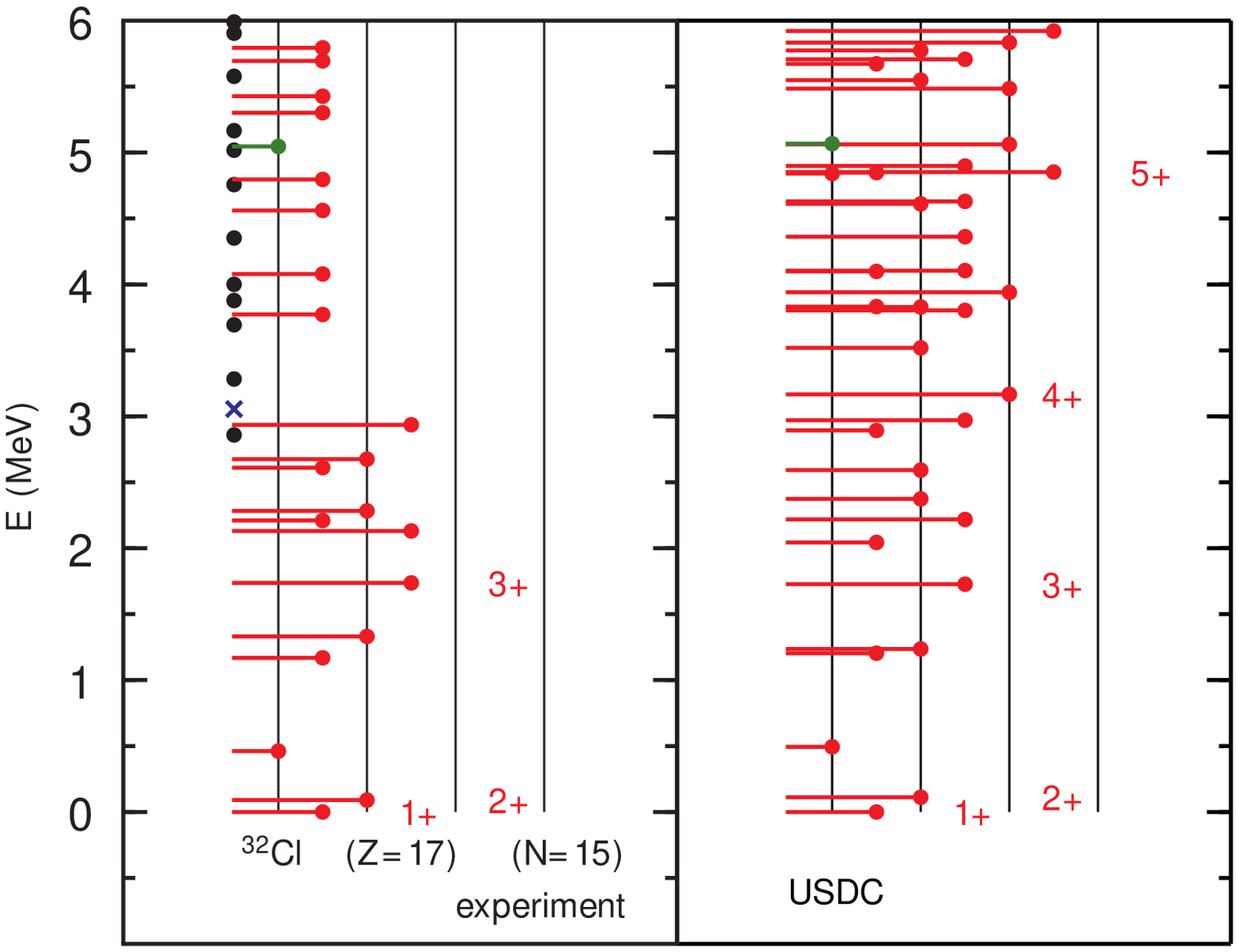}
\par\end{centering}
\caption{{\small{The left side of the figure shows results for $^{32}$P, whereas the right side shows $^{32}$Cl. Each side compares the position of experimental levels with predictions 
                 from a shell-model calculation with the USDC Hamiltonian. The Y axis gives the excitation energies. The length of the bars indicates the spin of the states, with each 
                 first spin level given in each part of the figure. The blue bar (or "x") is the first negative parity state, a 2$^-$ state in the case of $^{32}$P. 
                 The 0$^+$ T=2 level is shown by the green line.
}}}
\label{fig:p32}         
\end{figure*}

We start with understanding to what extent the USD-type Hamiltonians are able to describe the experimental spectra.
The energy levels for the mirror nucleus $^{32}$P are better known, so we use this nucleus for a first comparison of experiment and the USD-type hamiltonians. 
The experimental excitation energies for $^{32}$P and those obtained with the USDC Hamiltonian are shown in figure~\ref{fig:p32} (left part).
The results for other USD-type Hamiltonians are similar. The experimental energies are from~\cite{nndcA=32}. The lowest known
negative parity state, a 2$^{-}$ state at 3.3 MeV, is shown in blue. Other states with an assigned negative parity above that state are not shown.
These negative-parity states would be associated with 1$\hbar\omega$ excitations, being dominated by one nucleon moved from
the {\it sd} shell to the $pf$ shell, and are not within our model space.

There is a good match between experiment and theory up to 2.3 MeV. Two states with unassigned spin-parity at 2.3 and 2.6 MeV
are indicated by the black dots. These are only seen in one experiment, $^{31}$P(n,$\gamma$), and need to be confirmed.
Above that there is good agreement between experiment and theory for positive-parity states up to about 5 MeV.
The rms deviation between the theoretical and experimental excitation energies up to 5 MeV is about 150 keV.
This is true for all the USD-type Hamiltonians and is consistent with the rms deviations observed with these Hamiltonians in other nuclei
\cite{brown06,magilligan20}.

If there are pairs of experiment-theory levels with excitation energies that agree within 200 keV we consider this "good agreement".
There is a well established low-lying 0$^+$ T=1 state at 512.7~keV in agreement with theory, and the 0$^+$ T=2 at 5.072 MeV is also in agreement with theory.
Below 6 MeV there is only one more 0$^+$ T=1 state at 4.81~MeV in comparison with a tentative (0$^+$) state experimentally at 4.62 MeV.  
The location of these 0$^+$ T=1 states is important for the interpretation of isospin mixing with the 0$^+$ T=2 state \cite{signoracci11}.

Below the 0$^+$ T=2 state, there are seven 1$^+$ states in the calculation compared to eight observed experimentally. Starting around 5~MeV the
experimental level density becomes larger than theory. For example, there are two 2$^+$ states near 5 MeV that do not have a theoretical counterpart.
Thus, starting at about 5 MeV there are levels in experiment that are intruder states relative to the {\it sd}-shell model space.
These would be associated with 2$\hbar\omega$ excitations, being dominated by two nucleons moved from the {\it sd} shell to the $pf$ shell.
This means that a direct comparison of experimental and USD model-space states above about 5~MeV is no longer straight forward.

The USD Hamiltonians provide a good starting point for the wavefunctions. But the true Hamiltonian differs from the USD Hamiltonian by an rms average of 150~keV 
in the diagonal many-body matrix elements. We might assume that there are off-diagonal matrix elements of a similar size.
The reason comes from the assumption made about the USD-type Hamiltonians, i.e. that they apply only to the {\it sd} model-space
degrees of freedom, and that they are the same (except for some smooth mass scaling) for all states ascribed to the {\it sd} model space.

The implication of this is that matrix elements associated with observables such as $\beta$ decay, $\gamma$ decay, and spectroscopic factors will have 
uncertainties due to level mixing. For example, consider the $\beta$ decay of $^{32}$Ar to two 1$^+$ states in $^{32}$Cl that have a USD energy 
separation of $\Delta$ = 1 MeV with USD B(GT) values of 0.5 and 0.1. The experimental states $\mid 1>$ and $\mid 2>$ will be mixtures of the USD states
$\mid a>$ and $\mid b>$ with $\mid 1>\, = \alpha  \mid a> +\, \beta  \mid b>$ and $  \mid 2>\, = \beta \mid a> -\,  \alpha  \mid b>$. If mixing is obtained from a
two-dimensional matrix with an off-diagonal matrix element of $\delta  = \pm 150$ keV, the resulting error on the B(GT) is approximately 0.07 for both, with the
sum of 0.60 being exactly the same. Thus, the error on these quantities is not simply proportional to their values, but is the same for both,
0.07 in this example. The implication is that the theoretical error associated with observables is not proportional to the size
of the observable, but is rather closer to a constant value. This is consistent with the comparison between
experiment and theory made for a large number of observables in \cite{richter08}.

\subsubsection{Comparison of experiment and theory for $^{32}$Cl}

In the $\beta$ decay of $^{32}$Ar, seven 1$^+$ states plus the 2611~keV state discussed above are observed below the 0$^+$ T=2 state compared to the 8 expected from levels in $^{32}$P  (see figure~\ref{fig:p32}, right part). 
Up to about 3~MeV excitation energy, a one-to-one match of states in $^{32}$P and $^{32}$Cl exists in experiment and theory. Above that energy too many states observed in $^{32}$Cl
have still an unknown spin/parity and renders a further comparison difficult. It is reasonable to assume that the first intruder state is the negative parity state at around 3~MeV, 
as in the case of $^{32}$P.

\begin{figure}[hht]
\begin{centering}
\includegraphics[scale=0.37,angle=-90]{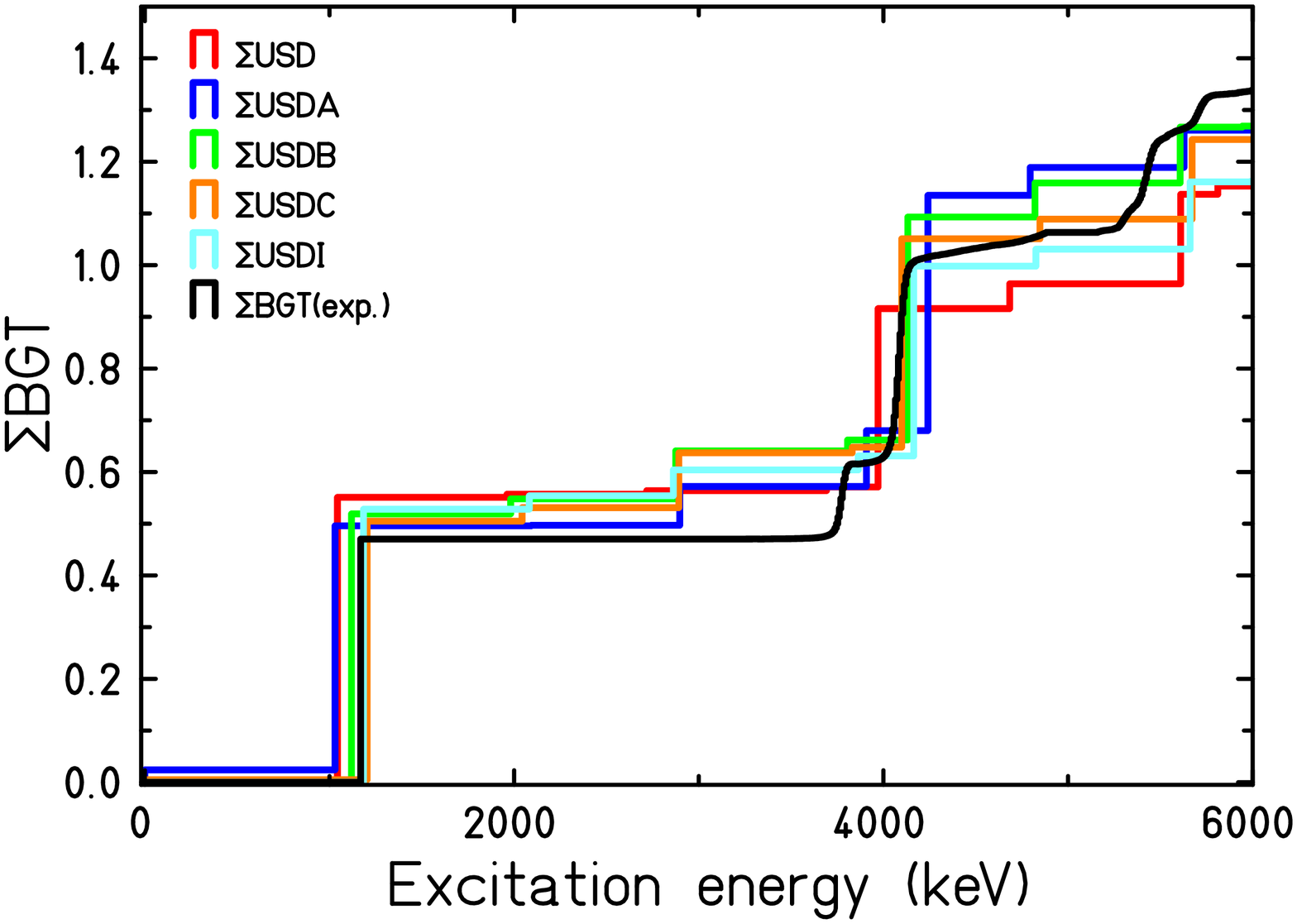}
\par\end{centering}
\caption{\small{The figure shows the evolution of the summed Gamow-Teller distribution for the $^{32}$Ar $\beta$ decay as a function of excitation energy in the lower-energy region on a linear scale. 
                 The different USD-type interactions
                 are compared to the experimental distribution from the full proton spectrum. The spread of the model predictions gives an indication of the model uncertainty.
}}
\label{fig:bgtlow}         
\end{figure}

The B(GT) values for our five Hamiltonians are compared to experiment in figure~\ref{fig:bgtlow} up to 6~MeV on a linear scale. 
We use a quenching factor of about 0.6 obtained from the ratios of the theoretical over the experimental half-life for the different interactions. 
All of the USD Hamiltonians give similar results and are in qualitative agreement with experiment, in particular for the
strong transitions at 1.168~MeV and 4.082~MeV followed by a gradual rise of the B(GT) sum up to 6 MeV. Above 5 MeV the calculated 1$^+$ states must mix with the intruder 
1$^+$ states (see the discussion for $^{32}$P). This mixing spreads the strength, but does not change the summed value.

The spread of the five calculations provides a measure of the theoretical uncertainty with the model assumptions of a universal {\it sd}-shell Hamiltonian. 
The experimental results generally lie within this spread. For the weak state near 2.2 MeV, the experimental B(GT) of 0.0020(4) is smaller than most of 
the calculations with the exception of USDA (0.0010). The weak state predicted near 2.7-2.8 MeV is not observed in experiment.
There is a suggested (1$^+$) at 2.611 MeV~\cite{nndcA=32}, which has a 1$^+$ mirror level in $^{32}$P. The experimental B(GT) to a state at this energy has an upper limit of B(GT) $<$ 0.0001.
Most of the calculations are much larger than this value with the smallest value coming from USD with B(GT) = 0.008. To obtain the small experimental value
one has to invoke state mixing beyond the model assumptions as discussed above.

\addtolength{\tabcolsep}{-5pt}
\begin{table*}[htt]
\caption{{\small{}\label{tab:usd}}
         The table compares the ratios of proton decay widths for three Gamow-Teller fed states and for the IAS for proton emission to the ground state over
         the emission to the first excited state of $^{31}$S. Two model calculations with the USDC and USDI hamiltonians~\cite{magilligan20} are compared
         to the experimental ratios of the proton branching ratios to the ground and first excited states. For the third state, no experimental branch to the first excited 
         state was observed. We therefore use as an upper limit the smallest branching ratio observed in the present work of 0.005\% with a 100\% error bar.}
\begin{center}
\normalsize{
\begin{tabular}{ccc|cc|cc}
\hline 
\hline 
\rule{0pt}{1.3em}
&\multicolumn{2}{c}{USDC} & \multicolumn{2}{c}{USDI} & \multicolumn{2}{c}{exp.} \\
& E (MeV) & $\Gamma_p$(g.s.) / $\Gamma_p$(ex. state) &  E (MeV) & $\Gamma_p$(g.s.) / $\Gamma_p$(ex. state) &  E (MeV) & BR(g.s.) / BR(ex. state) \\
&  3.8320 &  13.33 &  3.8640 &   9.41 &  3.7717 & 258(147)  \\
&  4.0990 &  11.73 &  4.1640 &   9.32 &  4.0816 &  22(1)    \\
&  4.8480 &  51.85 &  4.8260 &  58.33 &  4.7990 & $>$56(56)   \\
&  5.0670 &  47.86 &  5.0560 &  55.00 &  5.0463 &  77(2)    \\
[0.5em]\hline 
\hline 
\end{tabular}
}
\end{center}
\end{table*}

In figure~\ref{fig:bgt}, we show the full B(GT) distribution up to the $Q$ value limit. Evidently, the match between experiment and theoretical predictions 
is excellent up the highest states populated in the $\beta$ decay of $^{32}$Ar. The total B(GT) strength summed 
over all final states for all these Hamiltonians is about 7.0. Thus about 60\% of the Ikeda Gamow-Teller strength sum rule is observed in the $^{32}$Ar $\beta$ decay window.

We have also calculated the proton-decay and $\gamma$-decay widths for the 1$^+$ states with the methods used for $rp$ reaction rates in \cite{richter11}.
For the proton-decay width we calculate $\Gamma _{p} = {\rm C^{2}S} \,\, \Gamma _{sp}$, where C$^{2}$S are the spectroscopic factors obtained from the
{\it sd}-shell wavefunctions, and $\Gamma_{sp}$ are the single-particle proton-decay widths obtained from proton scattering from
a Woods-Saxon potential with the resonance energy constrained to the experimental value. The states that are unbound to proton decay to the ground state
and first excited state in $^{31}$S have proton-decay widths that are much larger than the $\gamma$-decay widths. This is consistent with the present 
observations. The proton-decay branching ratio for the decay to the ground state divided by the branching ratio to the first excited state are
compared between experiment and theory in table~\ref{tab:usd}. The largest disagreement is observed for the state near 3.8 MeV
that is relatively strong in the $^{32}$Ar $\beta$ decay. However, the calculated $\ell$=0 spectroscopic factors are very small.
For USDC they are 0.007 to the $^{31}$S 1/2$^+$ ground state and 0.0053 to the $^{31}$S 3/2$^+$ excited state. Such small spectroscopic factors
are very sensitive to the state mixing beyond the model assumptions.


\subsection{The Fermi strength to the IAS}

The feeding of the IAS in $^{32}$Cl can be determined in part from its decay by proton emission to the ground (proton peak 14) and the first excited (peak 6) states. Although the IAS 
is largely proton unbound and one would expect that it decays only by proton emission, Bhattacharya {\it et al.} have shown that it decays also by $\gamma$-ray emission to three states 
in the $\beta$-decay daughter nucleus $^{32}$Cl. We can clearly identify the first two $\gamma$ rays (labelled $j$ and $l$ in table~\ref{tab:ias_gam}), whereas we consider our evidence 
for the highest energy $\gamma$ ray only tentative (labelled $p$ in the same table). This $\gamma$ decay of the unbound IAS is favoured due to the isospin-forbidden character of
proton emission from the IAS. Only due to a most likely quite small T=1 impurity of the IAS, this proton emission can happen. The proton-emission branching ratio is determined
in our work to be 20.77(17)\%, whereas the $\gamma$-ray branches sum up to 1.86(18)\% yielding a total branching ratio to the IAS of 22.63(25)\%. All these values are in perfect 
agreement with those already given by Bhattacharya {\it et al.}~\cite{bhattacharya08}. In fact, this is not very astonishing, as they are averages from the present data with those 
from Bhattacharya {\it et al.} and from Bj{\"o}rnstad {\it et al.} and include in addition the total proton-emission branching ratio from Bhattacharya {\it et al}.

This super-allowed branching ratio allows us to determine the $ft$ value of the super-allowed decay branch to be 1514(35), where we used~\cite{nndcA=32} a half-life of 98(2)~ms, a $Q$ value of 
11134.3(18)~keV for the ground-state decay  and thus a $Q$ value to the IAS of 6088.0(18)~keV yielding f~= 3494.6(58). An electron-capture probability of $P_{EC}$~= 
0.067\%~\cite{firestone} was also included. To determine the corrected $\mathcal{F}t$ value (see e.g.~\cite{hardy20}), we use in addition the theoretical isospin-breaking 
correction $\delta_c$~=2.0(4)\% and the radiative correction $\delta_R$~= 1.145(41)\% from Bhattacharya {\it et al.} and obtain $\mathcal{F}t$~= 1501(35)~s. As this is a 
value for a T=2 transition, it can be compared to the value determined for the well-known 0$^+ \rightarrow$ 0$^+$, T=1 decays compiled regularly by Hardy and 
Towner~\cite{hardy20} divided by a factor of 2. The value from the  latest compilation is 3072.24(57)~s, which shows that the presently determined value is in 
agreement with expectations. 

The isospin non-conservation that is contained in the USDC and USDI Hamiltonians results in isospin mixing between the 0$^+$, T=2 state and 0$^+$, T=1 states.
The strongest mixing is with a state 226 (308) keV lower in energy with the USDC (USDI) hamiltonian.
This results in a splitting of the total Fermi strength of B(F)=4 into B(F)=3.87 (B(F)=3.92) for the T=2 state and 0.13 (0.07) for the T=1 state below the 
IAS for the USDC (USDI) interaction. The remaining Fermi strength is distributed over the higher-lying 0$^+$ states in the region of 6 to 10~MeV not observed in the present experiment.

The calculated proton-decay width of this 0$^+$, T=1 state is 36 keV (C$^2$S=0.042) with USDC and 37 keV (C$^2$S= 0.042).
A possible candidate for the 0$^+$, T=1 state below the IAS could be the experimentally observed state at 4779~keV. However, the branching ratio to this state is about a factor of 25 lower 
than predicted. Another candidate could be the state at 4561~keV. However, its  branching ratio is about a factor of 8 lower than predicted.

From the branching ratios for the IAS, we determine a ratio of the decay widths of R = $\Gamma_p$/$\Gamma_g$ = 11.2(11). The literature gives a total width of the IAS of $\Gamma$~= 
$\Gamma_p$ + $\Gamma_g$ = 20(5) eV~\cite{nndcA=32}. These values allow us to determine $\Gamma_g$ = $\Gamma$/(1+R) = 1.8(5). This compares well with the calculated value of
$\Gamma_g$~= 1.3~eV from both USDC and USDI. The experimental proton width is thus $\Gamma_p$~= 18.2(5) eV and has to be compared to the calculated values of 525~eV for USDC and 600~eV for USDI.
Since the proton width $\Gamma_p$ comes from isospin mixing, the conclusion is that the isospin mixing from the experiment is much smaller than in the
calculations, which will have consequences for the interpretation of the isobaric multiplet mass equation for A=32 \cite{signoracci11}. This conclusion is confirmed by the determination
of the experimental B(F) value calculated from the $ft$ value of the super-allowed decay branch to be 3.991(93) in agreement with the expected value of 4, but also with the theoretical values of 3.87/3.92.

\section{Conclusion}

We have presented a complete study of the $\beta$ decay of $^{32}$Ar performed at the identification station of SPIRAL1 with the silicon cube detection array and three
EXOGAM $\gamma$-ray detectors. The data obtained for $\beta$-delayed protons and $\gamma$ rays were compared to data from Bj{\"o}rnstad {\it et al.} and Bhattacharya {\it et al.}
and good agreement was obtained. This allowed us to average the available experimental data, determine feedings of excited states of $^{32}$Cl and extract $ft$ values and the
Gamow-Teller strength distribution. This distribution was compared to shell-model calculations using the USD, USDA, USDB, USDC, and USDI interactions and excellent agreement
was obtained. Finally, the Fermi strength was determined to be in agreement with the value expected for a super-allowed 0$^+ \rightarrow$ 0$^+$ decay. Although there might be two candidate
states for a 0$^+$, T=1 state, which could mix with the 0$^+$, T=2 isobaric analog state, we believe that these states are rather Gamow-Teller fed states.

\section*{Acknowledgment}

We express our gratitude to the EXOGAM collaboration for providing us with the germanium detectors. 
We thank the whole GANIL and, in particular, the accelerator staff for their support during the experiment. 
This work was partly funded by the Conseil r{\'e}gional de la Nouvelle Aqui\-taine and the EU through the Human
Capital and Mobility program. We acknowledge support from CICYT via contract FPA2007-62179.
BAB is supported by NSF via grant PHY-1811855.

\end{document}